\documentclass{article}
\usepackage{neurodist_preprint,times}
\usepackage[T1]{fontenc}
\usepackage[utf8]{inputenc}
\usepackage{amsmath,amssymb,booktabs,graphicx,microtype}
\usepackage{tikz}
\usetikzlibrary{arrows.meta,positioning,calc}
\usepackage{hyperref,url}
\usepackage{xurl}
\hypersetup{colorlinks=true,citecolor=blue!55!black,linkcolor=blue!55!black,urlcolor=blue!55!black,
  pdfauthor={Maryna Kapitonova and Tonio Ball},pdftitle={Distilling neural population geometry into Transformers and GRUs for motor control}}

\title{Distilling neural population geometry into\\Transformers and GRUs for motor control}
\author{Maryna Kapitonova$^{1,2}$\thanks{\raggedright Correspondence: \texttt{maryna.kapitonova@neuromentum.ai} and \texttt{tonio.ball@neuromentum.ai}.} \quad Tonio Ball$^{1,2}$\footnotemark[1]\\
$^1$NeuroMentum AI\\
$^2$Intelligent Machine-Brain Interfacing Technology (IMBIT),\\
University of Freiburg, Germany}

\begin{document}
\maketitle
\fancyhead{}
\pagestyle{plain}
\thispagestyle{plain}
\begin{abstract}
We introduce neuro-distillation (Neuro-Dist), a technique designed to directly transfer computationally useful neural population structure from biological brains into artificial controllers.
Using three macaque motor-cortical datasets as neuro-distillation teachers, we align controller internal states with cortical neural population geometry during training; controller deployment requires no neural data. We evaluate gated recurrent unit (GRU) networks and Transformers controlling a simulated arm.
Neuro-Dist achieves the lowest median error among five methods, including distillation from movement kinematics, in 10 of 18 comparisons across trained, student-held-out, and new target directions.
Seed-split aggregate statistical comparisons significantly favor Neuro-Dist over all six controls on trained and new targets, and five on student-held-out targets (Holm-corrected familywise $p<0.05$).
We furthermore find that Neuro-Dist can substantially improve controller robustness to sensory feedback noise.
Mechanistically, Neuro-Dist induced low tangling of trajectories in internal controller state space---a functionally important characteristic of motor cortex, demonstrating successful transfer of a known motor-cortical population geometry feature.
These findings establish neuro-distillation as a complement to behavioral distillation and point to a novel class of neuro-interfaces that tap biological neural population structure as a computational resource for improving the control of goal-directed behavior in robotics and embodied AI.
\end{abstract}

\section{Introduction}
\label{sec:introduction}

Behavioral control in embodied AI agents and robotic systems is typically learned from task objectives, demonstrations, or simulated experience. Examples of large-scale efforts include collecting demonstrations with 13 physical robots over 17 months \citep{brohan2022rt1} and training a manipulation policy with 131,072 parallel simulated robot instances over approximately 32 hours \citep{handa2022dextreme}. In contrast, mammalian behavioral control builds on neural circuitry shaped by hundreds of millions of years of vertebrate evolution across successive generations of global populations \citep{zador2019critique}. As a reference point for mammalian population scale, for example the contemporary abundance of the straw-coloured fruit bat alone---a single species with dexterous flight control---is estimated at roughly one billion individuals \citep{greenspoon2023biomass,omara2019acceleration}. Evolution shapes inherited circuit organization, while development and lifetime learning adapt these circuits to individuals and their environments \citep{zador2019critique}. Reproducing the breadth and adaptability of biological behavior in everyday physical environments remains an open challenge for artificial systems \citep{zador2019critique,tao2025dexwild}. This motivates investigating biological neural systems as a source of reusable computational structure that may be difficult to engineer \emph{a priori} and costly to empirically rediscover through training of artificial systems alone. As an example, neural population dynamics could supply training priors that transfer useful structure to artificial agents.

\begin{figure}[t]
\begin{center}
\includegraphics[width=\linewidth]{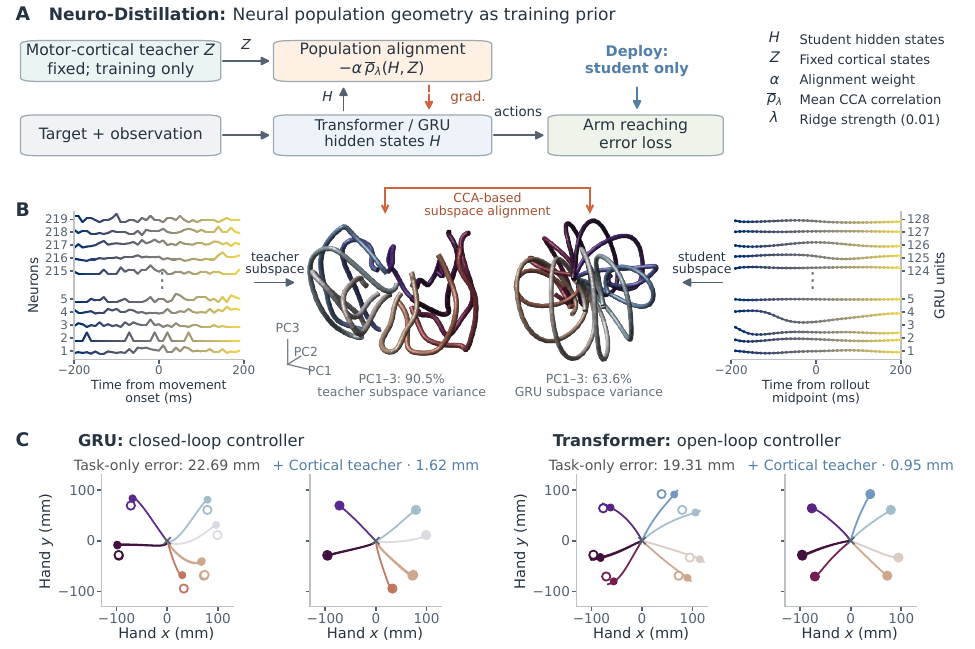}
\end{center}
\caption{\textbf{Neuro-distillation of neural population geometry into artificial controllers.} (A, B) A low-dimensional subspace is constructed from the neural teacher data recorded during reaching movements. PC1--PC3 denote the first three principal components (PCs). Two controllers (GRU, Transformer) are trained in a reaching task with a simulated arm. During neuro-distillation-guided training, a subspace of the student's hidden internal state is aligned to the teacher subspace via canonical correlation analysis (CCA) metrics. The students are deployed without further supervision and without requiring neural data at this stage. (C) Neuro-distilled controllers show more accurate target reaches; they also improved robustness against sensory feedback noise (see Figure~\ref{fig:sensorynoise}).}
\label{fig:framework}
\end{figure}

We investigate this possibility in motor control, where the mammalian central nervous system combines ancient spinal circuitry with more recently evolved cortical pathways \citep{jung2018ancient,rathelot2009subdivisions}. Two converging lines of research suggest that transferring biological representations to artificial motor controllers may be feasible. First, in the visual system, supervision derived from primate cortical recordings has improved image recognition \citep{federer2020statistical,safarani2021robust}, with representational alignment also improving visual-model robustness \citep{li2019brains,dapello2023aligning}. Second, studies of motor-cortical population geometry describe movement preparation and execution as low-dimensional manifolds \citep{churchland2012dynamics,russo2018untangled}. Such neural population geometry provides a conceptual bridge between biological and artificial systems \citep{chung2021geometry}; task-trained artificial recurrent networks can develop cortex-like dynamics \citep{sussillo2015muscle}. During reaching movements, a characteristic of population dynamics is \emph{low trajectory tangling}: nearby population states tend to have similar velocity vectors. This organization distinguishes motor-cortical activity from the muscle signals it helps generate and has been linked to noise robustness: Models trained to generate the same output using less-tangled internal trajectories tolerated more state noise \citep{russo2018untangled}. Primate motor cortex therefore offers a biologically motivated source of neural population structure for artificial controllers.

Here, we bring these lines together and ask: \emph{Can recorded motor-cortical population geometry serve as a training prior that improves artificial movement control?} We examine four subquestions:
\begin{enumerate}
\renewcommand{\labelenumi}{(\roman{enumi})}
\item Does Neuro-Dist improve accuracy on trained targets relative to task-only training and alternative teachers, including kinematics-based, engineered, stochastic, and shuffled controls?
\item Does Neuro-Dist also improve accuracy on targets not used during controller training?
\item Can Neuro-Dist increase controller robustness to sensory feedback noise?
\item Which characteristic of neural population geometry does Neuro-Dist transfer to students?
\end{enumerate}

To address these questions, we construct cortical teachers from primary motor-cortical spiking activity recorded in three macaques during reaching (Figure~\ref{fig:framework}). We train GRU and Transformer controllers and compare accuracy and generalization against task-only training and alternative teachers. Our results establish a proof of concept for neuro-distillation improving artificial motor control.

\section{Related work}
\label{sec:related}

\paragraph{Neural data as a teaching signal.}
Recorded activity may guide learning through example weighting, response prediction, representation matching, or evaluative feedback.
\citet{fong2018guide} weighted image-classification training errors with human functional magnetic resonance imaging (fMRI), needing no neural measurements at inference.
Brain-mediated transfer improves downstream label estimation by mapping audiovisual features through fMRI-learned representations \citep{nishida2020brain}.
\citet{mcclure2016representational} transferred representational distances between artificial networks and proposed extensions to measured brain representations.
Early electroencephalography (EEG)-guided image embeddings and joint image--EEG learning \citep{spampinato2017mind,palazzo2021decoding} have more disputed interpretations because of temporal correlations in block-designed data \citep{li2021perils,palazzo2020block}.
These approaches target perceptual representations and predictions; we supervise motor controllers based on motor-cortical single-unit data.

\paragraph{Visual representation learning.}
Neural regularization may transfer stimulus relationships and individual response targets.
\citet{li2019brains} improved image-model robustness using similarities from a mouse-visual-cortex predictor; \citet{federer2020statistical} improved recognition and corrupted-label tolerance using macaque response statistics.
Joint classification and macaque primary visual cortex response prediction improves image-distortion robustness \citep{safarani2021robust}; inferior-temporal-cortex alignment improves human behavioral agreement and adversarial robustness \citep{dapello2023aligning}.
A direct precursor, \citet{pirlot2022cca}, uses deep-CCA regularization to align convolutional neural network representations with monkey visual responses, improving classification and robustness to tested adversarial perturbations.
Human visual teachers include ReAlnet's EEG-response prediction, improving neural and behavioral alignment \citep{lu2026realnet}; fMRI-derived predictors and category-manifold guidance, improving robustness with area-dependent benefits \citep{shao2025robustness}; and fMRI--image structural alignment, improving visual-category generalization \citep{chen2026mental}.
In our study, we extend these concepts from the visual domain to upper-extremity motor control.

\paragraph{Speech, language, and reasoning.}
For sequential inputs, \citet{schwartz2019brain} fine-tuned Bidirectional Encoder Representations from Transformers (BERT) to predict fMRI and magnetoencephalography responses, preserving downstream natural language processing (NLP) performance.
Brain-response supervision improves speech recognition \citep{li2024speech} and, through fMRI brain-tuning, semantic speech tasks \citep{moussa2025braintuning}.
Joint cross-participant prediction extends brain-tuning benefits to neural prediction and semantic tasks \citep{moussa2025multibrain}; fMRI-tuned auditory networks improve several downstream auditory tasks \citep{freteault2025auditory}; bilingual fMRI supervision improves multilingual NLP \citep{negi2025multilingual}.
Time-resolved electrocorticography can supervise speech encoders \citep{zhang2026temporal}; fMRI-guided parameter fine-tuning improves evaluated deductive reasoning \citep{xiao2026reasoning}.
\citet{aquino2025brain2model} have reported improved memory-based decision learning by contrastively aligning internal network representations with human single-neuron embeddings.

\paragraph{Neural feedback and embodied motor learning.}
EEG error-related signals can teach robot control policies \citep{iturrate2015teaching,kim2017intrinsic}; decoded evaluative feedback accelerates robot navigation learning \citep{akinola2020brain}.
Here brain signals evaluate actions during deployment and provide a reward signal, e.g., for reinforcement-learning-based optimization; we distill internal controller states directly.
Behaviorally trained virtual animals assess post-training neural correspondence \citep{aldarondo2024rodent}; neurocybernetic modeling proposes distillation from neural-data-trained models, then embodied refinement \citep{park2026neurocybernetic}.
The closest embodied comparison, the neurally constrained nuSim variant of $\mu$Sim \citep{almani2025musim}, fits selected recurrent neural network (RNN) units to recorded single-unit firing rates, improving neural reconstruction at unseen cycling speeds. Its aim is to model motor cortex faithfully; controller performance or robustness benefits were not established. Our CCA-based population supervision avoids neuron-to-unit assignments and targets improved artificial control.

\setcounter{topnumber}{1}
\setcounter{totalnumber}{1}
\begin{table}[t]
\caption{\textbf{Reaching accuracy across target sets.} Median $\pm$ bootstrap standard error (SE) of final-window coordinate mean squared error (MSE; mm$^2$; 100 seeds), at training-selected strengths fixed across target sets. Gaussian and Poisson columns continue in Supplementary Table~\ref{tab:stochastic-controls}. Bold values mark row minima. M1--M3: Monkeys~1--3; Transf.: Transformer.}
\label{tab:main}
\begin{center}
\begingroup
\setlength{\tabcolsep}{2pt}
\begin{tabular}{llrrrrr}
\toprule
Student & Monkey & Task-only & Ring & Neuro-Dist & Behavior & Shuffled \\
\midrule
\multicolumn{7}{l}{\emph{A. Trained targets (28)}} \\
GRU & M1 & $58.42\pm3.48$ & $54.95\pm4.54$ & $\mathbf{33.66}\pm3.24$ & $38.57\pm3.77$ & $50.79\pm2.93$ \\
GRU & M2 & $60.72\pm4.79$ & $53.94\pm3.33$ & $\mathbf{26.83}\pm3.62$ & $38.60\pm2.69$ & $50.13\pm3.57$ \\
GRU & M3 & $55.59\pm3.71$ & $52.00\pm2.59$ & $48.41\pm3.19$ & $\mathbf{46.56}\pm2.55$ & $59.10\pm4.08$ \\
\addlinespace[2pt]
Transf. & M1 & $24.29\pm2.34$ & $25.69\pm1.50$ & $12.24\pm3.38$ & $\mathbf{9.77}\pm1.37$ & $39.40\pm5.08$ \\
Transf. & M2 & $31.40\pm6.18$ & $12.53\pm2.40$ & $\mathbf{8.63}\pm1.86$ & $15.79\pm3.04$ & $28.48\pm3.57$ \\
Transf. & M3 & $28.16\pm6.20$ & $20.58\pm2.55$ & $\mathbf{15.90}\pm2.98$ & $16.75\pm1.46$ & $30.08\pm2.19$ \\
\midrule
\multicolumn{7}{l}{\emph{B. Generalization to student-held-out targets (8)}} \\
GRU & M1 & $62.05\pm2.72$ & $59.05\pm5.90$ & $42.98\pm3.96$ & $\mathbf{42.20}\pm5.33$ & $56.55\pm5.13$ \\
GRU & M2 & $64.80\pm5.31$ & $60.09\pm5.02$ & $\mathbf{31.97}\pm3.88$ & $41.36\pm2.22$ & $55.54\pm3.79$ \\
GRU & M3 & $59.62\pm3.67$ & $53.91\pm3.45$ & $\mathbf{53.05}\pm2.97$ & $53.39\pm2.31$ & $61.77\pm3.73$ \\
\addlinespace[2pt]
Transf. & M1 & $27.66\pm2.36$ & $31.49\pm2.96$ & $18.32\pm3.75$ & $\mathbf{12.60}\pm1.78$ & $48.77\pm4.34$ \\
Transf. & M2 & $34.13\pm8.45$ & $13.89\pm2.27$ & $\mathbf{9.49}\pm1.67$ & $16.61\pm2.78$ & $28.28\pm3.65$ \\
Transf. & M3 & $30.38\pm9.74$ & $20.42\pm2.57$ & $17.99\pm2.96$ & $\mathbf{16.60}\pm1.67$ & $35.93\pm2.57$ \\
\midrule
\multicolumn{7}{l}{\emph{C. Generalization to new target directions (8)}} \\
GRU & M1 & $79.82\pm3.09$ & $76.94\pm3.64$ & $62.44\pm4.84$ & $\mathbf{56.81}\pm6.10$ & $95.63\pm8.04$ \\
GRU & M2 & $58.80\pm3.51$ & $53.69\pm4.20$ & $\mathbf{29.57}\pm3.56$ & $38.90\pm3.16$ & $50.68\pm3.34$ \\
GRU & M3 & $55.24\pm3.03$ & $50.61\pm3.55$ & $47.82\pm3.25$ & $\mathbf{47.18}\pm3.95$ & $53.63\pm3.63$ \\
\addlinespace[2pt]
Transf. & M1 & $91.23\pm17.11$ & $106.44\pm11.76$ & $66.61\pm15.42$ & $\mathbf{65.72}\pm7.55$ & $125.30\pm11.68$ \\
Transf. & M2 & $30.23\pm5.54$ & $14.86\pm2.53$ & $\mathbf{10.28}\pm1.54$ & $18.29\pm2.79$ & $31.19\pm3.74$ \\
Transf. & M3 & $27.23\pm5.38$ & $22.44\pm3.19$ & $\mathbf{16.26}\pm2.12$ & $18.07\pm1.72$ & $35.22\pm2.79$ \\
\midrule
\multicolumn{2}{l}{Sum of wins} & 0 & 0 & \textbf{10} & \textbf{8} & 0 \\
\bottomrule
\end{tabular}

\endgroup
\end{center}
\end{table}

\begin{figure}[t]
\begin{center}
\includegraphics[width=\linewidth]{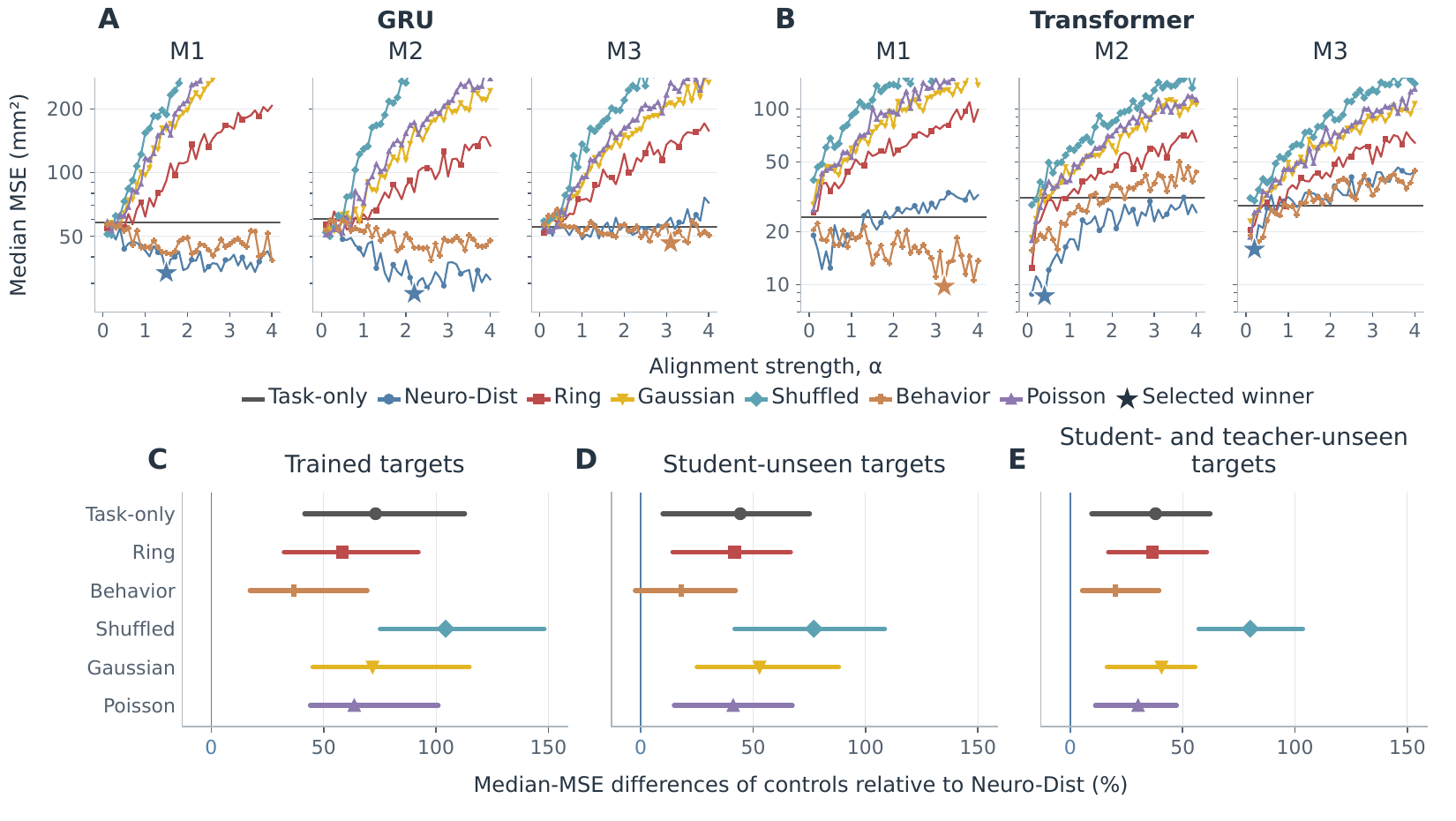}
\end{center}
\caption{\textbf{Alignment response and statistical evaluation.} \textbf{A,B:} GRU and Transformer trained-target median MSE (100 seeds); stars mark winners across all seven methods at training-selected strengths. \textbf{C--E:} Control median-MSE differences relative to Neuro-Dist, from geometric-average ratios across both architectures and three animals (50 evaluation seeds). Positive values indicate increased control errors; 0\% matches Neuro-Dist performance. Bars show pointwise 95\% confidence intervals from bootstrap resampling. Statistical details: Appendix~\ref{app:accuracy-statistics}. In aggregate comparisons, Neuro-Dist significantly outperforms all six controls on trained and new targets and five on student-held-out targets, with no control significantly outperforming Neuro-Dist (Holm-corrected $p<0.05$).}
\label{fig:dose}
\label{fig:accuracy-statistics}
\end{figure}

\begin{figure}[t]
\begin{center}
\includegraphics[width=\linewidth]{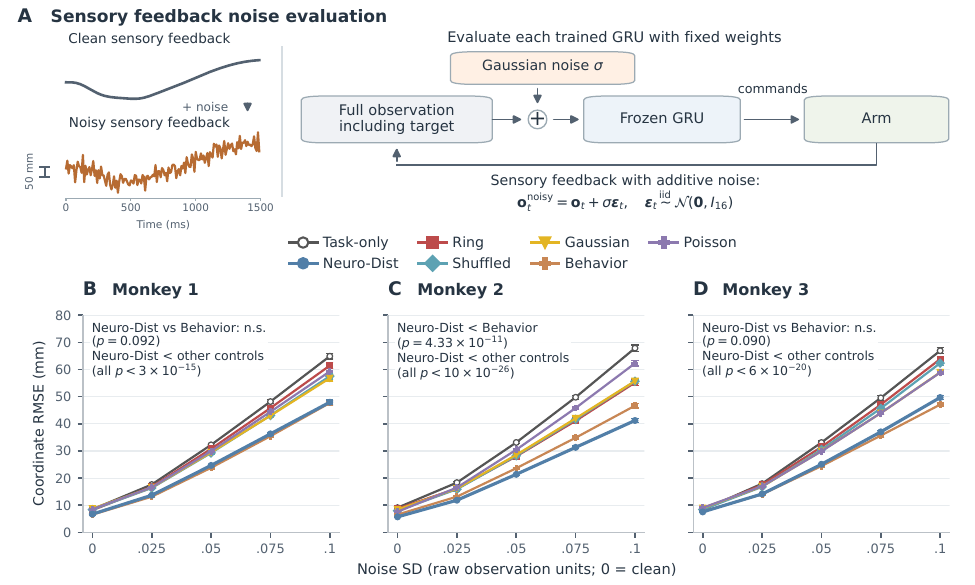}
\end{center}
\caption{\textbf{Neuro-Dist improves GRU robustness to sensory feedback noise.} (A) We perturb all 16 observation channels of frozen GRUs trained without noise. Gaussian noise is independent across channels and 10-ms steps: $\mathbf{o}_t$ and $\mathbf{o}^{\mathrm{noisy}}_t$ denote clean and noisy inputs, $\sigma$ the noise standard deviation (SD) in channel units, and $\boldsymbol{\epsilon}_t$ a standard-normal vector. The separate 1.5-s illustration uses a fixed-dose controller ($\sigma=25$\,mm for positions). (B--D) Mean coordinate RMSE $\pm$ SEM (standard error of the mean) across 100 seeds, each averaging three realizations per positive amplitude. Strengths are selected on 28 clean training targets; 400-ms evaluations score eight student-held-out targets. Annotations: paired signed-rank tests of seedwise mean error ratios, Holm-corrected across 18 comparisons. ``$<$'': lower error; ``n.s.'': nonsignificant. Ratio definitions and confidence intervals: Supplementary Figure~\ref{fig:grunoiseheldoutcomparisons}. Neuro-Dist and Behavior together have the lowest mean errors at every noise level. Their primary comparison favors Neuro-Dist in Monkey~2, with no significant difference in Monkeys~1 and~3; no control significantly outperforms Neuro-Dist in any primary comparison.}
\label{fig:sensorynoise}
\end{figure}

\begin{figure}[t]
\begin{center}
\includegraphics[width=\linewidth]{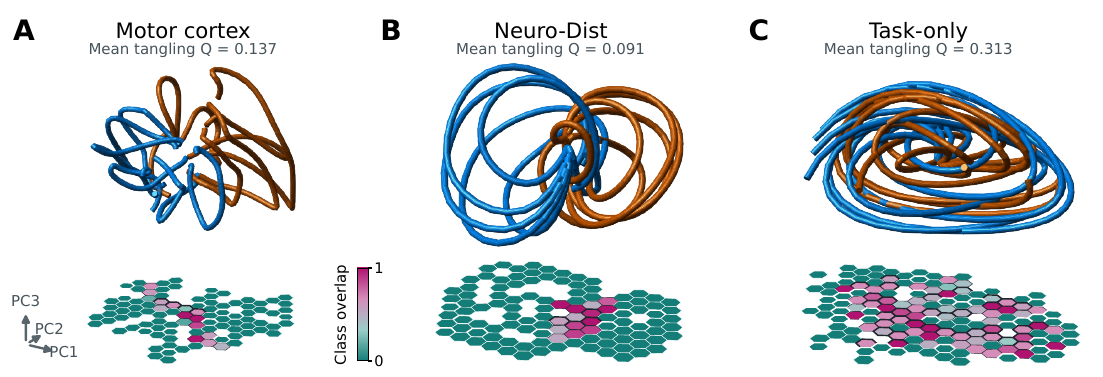}
\end{center}
\caption{\textbf{State-space tangling of neural data and controllers\@.} \textbf{(A)} Example of subspace projection of Monkey~2 cortical activity from one contributing session and \textbf{(B)} of a Neuro-Dist GRU ($\alpha=2.2$, seed~45); \textbf{(C)} its matched task-only GRU\@. Principal component analysis (PCA) projections show 12 representative reaching conditions; hexagons measure projected direction-class overlap. Tangling $Q$ is low in both the motor cortical recordings and the neuro-distilled controller, but substantially higher in the task-only controller trained without any distillation.}
\label{fig:tangling}
\end{figure}

\begin{figure}[t]
\begin{center}
\includegraphics[width=\linewidth]{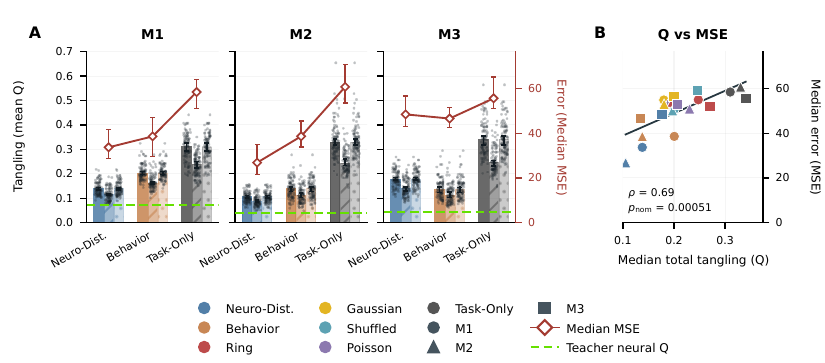}
\end{center}
\caption{\textbf{Neuro-distillation reduces GRU state-space trajectory tangling.} \textbf{A,} Bars: median total, within-reach and between-reach $Q$ (left to right); dots: individual models. Red diamonds: median training MSE (mm$^2$); whiskers: pointwise 95\% bootstrap intervals across 100 seeds. Green dashed lines: cortical references. Neuro-Dist has the lowest median tangling and training error in M1--M2; Behavior has both minima in M3 (Table~\ref{tab:main}A, GRU rows). \textbf{B,} Lower tangling correlates with lower reaching error across 21 dataset--method conditions. See Appendix~\ref{app:tangling} for analysis and display details.}
\label{fig:gru-tangling}
\end{figure}

\section{Methods}
\label{sec:methods}

\paragraph{Population geometry as a training prior.}
Neuro-distillation combines conventional reaching supervision with an \emph{a priori} cortical-population alignment objective. Neural teachers use publicly available, condition-averaged motor-cortical data from three macaques performing reaching tasks: Monkey~1 from \citet{pei2021nlb} and Monkeys~2--3 from \citet{odoherty2017reaching}. We project 12 cortical PCs onto a rank-four subspace predictive of recorded hand position and velocity, retaining 12 coordinates. All 36 reaching conditions enter teacher construction, including those withheld from student training. Regularized CCA rewards correlated linear projections of student and fixed teacher activity across matched conditions and time points \citep{hotelling1936relations,andrew2013dcca}. This encourages shared linear population geometry without neuron-to-unit assignments or requiring identical full-state distances or dynamics. Recordings are needed only during training; Appendix~\ref{app:detailed-methods} gives implementation details and Appendix~\ref{app:protocols} the protocols.

\paragraph{Task, students, and optimization.}
A closed-loop GRU with updated sensory feedback and an open-loop Transformer conditioned on initial observations and target information generate commands for MotorNet's differentiable two-joint, six-muscle arm \citep{codol2024motornet}. Students minimize
\begin{equation}
 \mathcal L_i=\mathcal L_{\rm reach}-\alpha\min(1,i/60)\overline{\rho}_\lambda(H,Z),
 \label{eq:main-objective}
\end{equation}
where $\mathcal L_{\rm reach}$ averages squared fingertip-to-target error across coordinates, the final 15 frames, and training conditions, normalized by squared target radius. The CCA score $\overline{\rho}_\lambda$ averages 12 regularized canonical correlations between student states $H$ and teacher states $Z$ (Suppl.\ equation~\ref{eq:cca}); $\alpha$ is the CCA alignment strength. $\alpha=0$ gives task-only training.

\paragraph{Evaluation.}
For each animal, architecture, and aligned teacher in Table~\ref{tab:main}, we test 40 strengths ($\alpha=0.1$--$4.0$), each with 100 seeds. Accuracy is the median final-15-frame coordinate MSE (mm$^2$) across all seeds, including high-error fits. These 100 seeds enabled robust median-based effect estimates, paired bootstrap confidence intervals, and Holm correction for multiple comparisons, with separate 50-seed groups for strength selection and evaluation (Appendix~\ref{app:accuracy-statistics}).

\paragraph{Controls.}
We compare Neuro-Dist to a comprehensive set of controls, each of them designed to probe a different competing distillation strategy or possible source of benefit from teacher supervision:
\begingroup
\begin{itemize}
\item \textbf{Task-only:} establishes performance without distillation: reaching loss alone ($\alpha=0$).
\item \textbf{Behavior:} tests supervision from recorded movement: four PCs of hand position and velocity at five temporal offsets, embedded in 12 coordinates.
\item \textbf{Ring:} an engineered directional prior motivated by directional organization in primary motor cortex data (see Appendix~\ref{app:directional-ring}).
\item \textbf{Gaussian:} controls for population variance structure: smooth random signals matching the normalized cortical covariance eigenvalue spectrum.
\item \textbf{Poisson:} tests supervision from synthetic spiking: animal-specific fitted rate distributions with smooth modulation, trial averaging, and reduction to four PCs.
\item \textbf{Shuffled:} tests the importance of correct condition correspondence: whole cortical trajectories are reassigned to different reaches, preserving time order.
\end{itemize}
\endgroup

\paragraph{Sensory feedback noise and trajectory geometry.}
Analyses of feedback noise and internal tangling are naturally suited to the closed-loop, recurrent GRUs; our open-loop Transformer neither receives sensory feedback nor evolves a recurrent hidden state. Independent Gaussian noise was used to perturb all feedback channels, including target information; errors use final-window coordinate root mean squared error (RMSE; Appendix~\ref{app:gruselectednoise}). Trajectory tangling quantifies velocity disagreement between nearby full-dimensional hidden states \citep{russo2018untangled}, within and between reaches (Appendix~\ref{app:tangling}). We define $Q_i=\max_j\|\dot{x}_i-\dot{x}_j\|_2^2/(\|x_i-x_j\|_2^2+0.1s_x^2)$, where $x_i$ is a hidden state, $\dot{x}_i$ its temporal derivative, and $s_x^2$ the total state variance; $j$ ranges over reference states. Reported $Q$ averages $Q_i$ over evaluated states; lower values indicate less tangling. Tangling is neither directly optimized during training nor used to select distillation strengths: it tests for a population-geometric feature beyond the alignment objective.

\section{Results}
\label{sec:results}

\paragraph{Does Neuro-Dist improve accuracy on trained targets relative to task-only training and alternative teachers?}
\label{sec:mainresults}
\label{sec:controls}

We evaluated GRU and Transformer controllers with teachers from three monkeys and 100 seeds per setting. Table~\ref{tab:main} reports median framewise coordinate MSE over the final 15 frames, retaining all seeds. Each aligned method uses a training-selected strength minimizing median training MSE over the same 40 positive doses; task-only by definition uses zero alignment.
On the 28 trained targets, Neuro-Dist has lower median MSE than task-only training in all six animal--architecture comparisons, with reductions of 12.9--72.5\% (Table~\ref{tab:main}A). Figure~\ref{fig:dose} shows the corresponding dose responses.

In a separate seed-split analysis, Neuro-Dist significantly outperforms all six controls on trained targets (tests based on paired seed-block bootstrap CIs; Holm-corrected familywise $p<0.05$ across 18 aggregate comparisons; Appendix~\ref{app:accuracy-statistics}). In Table~\ref{tab:main}, Neuro-Dist has the lowest median in four rows and Behavior in two; both have lower medians than Shuffled, and Neuro-Dist than Ring throughout. Gaussian and Poisson are additional controls with 40-dose coverage (Figure~\ref{fig:dose}; Supplementary Table~\ref{tab:stochastic-controls}).

\paragraph{Does Neuro-Dist also improve accuracy on targets not used during controller training?}
\label{sec:generalization}

We evaluated the same 3,000 selected controllers from the five methods in Table~\ref{tab:main} on two additional target sets, without retuning alignment strengths or retraining students.
\emph{(1) Generalization to student-held-out targets.}
Eight of the original 36 reaching conditions were withheld from student optimization but included in teacher construction. Neuro-Dist has lower median MSE than task-only training in all six comparisons, with reductions of 11.0--72.2\% (Table~\ref{tab:main}B). Supplementary Figure~\ref{fig:dosegeneralization1} shows the dose responses.
\emph{(2) Generalization to new target directions.}
To test generalization to conditions unseen by both teacher and student, we also evaluated eight directions per animal and seed outside the original 36-condition set used by teacher construction and student training. Neuro-Dist has lower median MSE than task-only training in all six comparisons, with reductions of 13.4--66.0\% (Table~\ref{tab:main}C) (Appendix~\ref{app:noveltargets}).

Corrected seed-split aggregate comparisons favor Neuro-Dist over all six controls on new targets and five on student-held-out targets; no statistically significant difference is detected against Behavior on the latter. Descriptively, Neuro-Dist and Behavior each have the lowest median in three rows per regime (10 versus 8 overall; Table~\ref{tab:main}); both outperform Shuffled, and Neuro-Dist outperforms Ring throughout. No control significantly outperforms Neuro-Dist in any aggregate comparison.

\paragraph{Can Neuro-Dist increase controller robustness to sensory feedback noise?}
\label{sec:robustness}

We tested 2,100 GRUs under white sensory feedback noise, selecting alignment strengths from clean training (Figure~\ref{fig:sensorynoise}). On student-held-out targets, Neuro-Dist lowers mean error versus task-only at every amplitude in all three animals; corrected tests also favor it over Ring, Shuffled, Gaussian, and Poisson. Primary comparisons with Behavior favor Neuro-Dist in Monkey~2, with no statistically significant difference detected in Monkeys~1 and~3. Sensory feedback noise raises mean RMSE less for Neuro-Dist than task-only at every positive amplitude in all animals (Appendix~\ref{app:gruselectednoise}). Thus, cortical supervision can improve GRU robustness to sensory feedback noise without noise-augmented training.

\paragraph{Which feature of neural population geometry does Neuro-Dist transfer to the students?}
\label{sec:tangling}

To address this question, we focused on low trajectory tangling, an established hallmark of motor-cortical population dynamics \citep{russo2018untangled}. We tested whether cortical alignment induces this feature in the student by measuring velocity disagreement between nearby hidden states. Tangling is neither a training penalty nor a dose-selection criterion. We analyzed all 128 hidden coordinates in 2,100 GRUs, with seven methods, three datasets, and 100 seeds; strengths minimize median training MSE over each aligned method's complete 40-dose grid. Figure~\ref{fig:gru-tangling} summarizes each model's mean $Q$ over all 40 frames of its 28 trained reaches, using all 36 conditions as references (Appendix~\ref{app:tangling}).

Neuro-Dist reduces median total $Q$ relative to task-only training by 55.5\%, 68.1\%, and 48.1\% in Monkeys~1--3, respectively. All five alternative teachers also reduce median tangling relative to task-only. Notably, Neuro-Dist has the lowest median total $Q$ among the seven methods in Monkeys~1 and~2---the same monkeys where Neuro-Dist substantially reduced reaching errors: tangling $Q$ was 0.138 and 0.105, versus 0.201 and 0.138 for Behavior. In Monkey~3, Behavior's $Q$ (and reaching error) is lower (0.134 versus 0.177), while Neuro-Dist remains below the other five controls. These Neuro-Dist--Behavior relationships also hold for within-reach and between-reach tangling. Crucially, and confirming the observation in these examples, across all 21 animal--method summaries, median tangling is positively and significantly associated with median training MSE (Spearman $\rho=0.692$, nominal $p=5.07\times10^{-4}$). Together with the biological importance of low tangling for robust motor control, these findings suggest transfer of a low-tangling state as a mechanism of the neuro-distillation success in achieving robust movement control also in artificial controllers.

The Neuro-Dist--Behavior ordering persists on student-held-out reaches and new-target evaluations, with all seven methods included in both analyses. Supplementary Figure~\ref{fig:gru-tangling-supplement} reports these comparisons and decompositions of velocity disagreement and state separation.
Figure~\ref{fig:tangling} illustrates the geometry using Monkey~2 as an example; the population comparisons above use all seeds and full hidden states.

\section{Discussion, limitations and conclusions}
\label{sec:discussion}

In the present study we show a proof of concept for neuro-distillation, a technique designed to directly transfer computationally useful neural population structure from biological brains into artificial controllers. Using three macaque intracortical datasets as neuro-distillation teachers, we successfully align the internal states of GRU and Transformer controllers with primary motor cortex population geometry extracted from single-unit spiking data.

We selected CCA-based subspace alignment for neuro-distillation \emph{a priori}, at the start of the project.
This choice was motivated by evidence that CCA can reveal shared, behaviorally relevant structure across neuronal populations, and is supported by successful applications of CCA-based alignment in brain-guided learning and model distillation \citep{wang2015privileged,pirlot2022cca,fan2025maximizing}.
During reaching, heterogeneous single-neuron responses collectively form organized, low-dimensional population trajectories \citep{churchland2012dynamics}.
Using CCA, \citet{gallego2020stable} revealed stable motor-cortical latent dynamics across recording sessions despite changes in the recorded neuronal sample, supporting reliable decoding of movement over time.
\citet{safaie2023preserved} subsequently used CCA to demonstrate preserved latent dynamics across animals performing similar behaviors, enabling movement decoding across individuals.
Together, these findings provided a physiological rationale for targeting shared linear population structure when aligning cortical activity with artificial controller states.
CCA accommodates different population sizes without requiring correspondence between individual neurons and artificial units. The current study shows that CCA-based neural supervision can be extended to the motor control domain, by aligning controller states directly with behavior-predictive neural teacher subspaces. Our findings demonstrate that this alignment concept improved reaching accuracy (Table~\ref{tab:main}) and increased robustness to sensory feedback noise in the recurrent controllers (Figure~\ref{fig:sensorynoise}B--D).
Interstingly, neuro-distillation based on trial-shuffled data was among the least accurate teachers (Table~\ref{tab:main}). This highlights that cortical statistics alone cannot explain the beneficial effects of neuro-distillation. 

Neuro-distillation reduced trajectory tangling in GRU population activity (Figure~\ref{fig:gru-tangling}A), despite tangling being neither an explicit training objective nor a criterion for model selection.
Low tangling characterizes motor-cortical activity, and recurrent-network experiments have demonstrated its contribution to noise robustness \citep{russo2018untangled}. Thus, our findings replicate this neurobiologically documented association between low tangling and improved reaching movements in our distilled controllers, across animal--method groups (Figure~\ref{fig:gru-tangling}B; Appendix~\ref{app:tangling}). These observations support a functional role of reduced tangling in the observed distillation control benefits.
However, our analyses are correlative and do not yet establish exactly to what extent other neural population features may have contributed to the observed control benefits.
An intriguing possibility therefore is that reduced tangling reflects only a facet of a broader transfer of computationally useful population structure, including features that our analyses may not yet have identified.
Candidates for further investigation include the rotational organization of movement-related population trajectories \citep{churchland2012dynamics} and their movement-dependent arrangement in neural state space \citep{sabatini2024reorientation}, which may provide a structured---and possibly distillable---basis for generating diverse motor commands. Together, these considerations also raise the question of whether cortical supervision may convey useful structure beyond that provided by any alternative, neuro-agnostic teacher such as the stochastic, behavior-based, and engineered teachers tested here. Future studies could probe these possibilities by selectively preserving or disrupting candidate features in neural teachers and, alongside matched controls, testing their causal contribution to controller accuracy and robustness.
Establishing which features account for the functional value of neuro-distillation would guide its development and could also offer a new paradigm to generate testable hypotheses to advance our mechanistic understanding of biological motor control.

Beyond the aspects discussed so far, the present proof of concept of neuro-distillation from biological neural systems to artificial controllers opens a broad research agenda: key questions are which aspects of biological population computation can improve motor control and goal-directed behavior in robotic systems and embodied AI agents, and what roles the teacher and student tasks and the kinds of effectors used (e.g., their biomechanical realism) play. Another crucial direction is translation from simulations, as studied here, to real-world settings, to investigate whether neuro-distillation can bring significant, relevant and lasting advantages in real-world applications.

Such applications range from dexterous manipulation and articulate speech production to whole-body coordination or expressive facial movements in humanoids, and the control of walking, flying, or swimming robots.
Benefits could encompass accuracy and robustness, but also learning efficiency, adaptation, and generalization to unfamiliar bodies or environments.
Different cortical areas, combinations of areas, species, and behavioral repertoires may offer useful structure for different computational demands.
Comparing these sources could distinguish broadly reusable principles from specializations associated with particular behaviors or embodiments.
At single-neuron resolution, technologies such as Neuropixels probes \citep{trautmann2025neuropixels}, wireless microelectrode arrays \citep{schwarz2014wireless}, and miniaturized calcium-imaging microscopes \citep{zhang2024mesoscope} provide complementary ways to sample population activity during behavior.
These approaches could help establish how population coverage, temporal resolution, and access to natural behavior influence the value of neural teachers.
Equally central is the design of potentially novel, optimized distillation objectives that preserve and allow transfer of the relevant aspects of population geometry, temporal organization, or coordination between neural processes.
Together, these topics and questions outline a research field on neuro-distillation as a new form of neuro-interfacing, linking the neurobiology of neural population computations and machine learning, with the aim to transfer useful computational structure to guide the computations learned by artificial controllers across tasks and embodiments.

In this study, we used publicly available macaque datasets; human neural data can be expected to offer important perspectives beyond animal recordings relevant to neuro-distillation in our sense, for example for dexterous hand and finger control or speech production. Such opportunities, however, would also come with new ethical questions, such as those regarding data privacy, which we advocate addressing as a crucial, integral research component right from the beginning \citep{unesco2025neurotechnology,kapitonova2022privacy}.

\label{maintextend}
\setcounter{topnumber}{2}
\setcounter{totalnumber}{3}

\clearpage
\appendix
\section*{Supplementary Material}
\section{Detailed methods}
\label{app:detailed-methods}

\paragraph{Neural teachers.}
\label{app:teachers}
We use frozen condition-averaged motor-cortical teachers from three macaques: Monkey~1 (MC\_Maze, distributed through the Neural Latents Benchmark; \citealp{pei2021nlb}) and Monkeys~2 and~3 (\citealp{odoherty2017reaching}). Each teacher contains 36 conditions and 40 time bins. Preprocessing uses 10-ms bins, Gaussian smoothing with a 30-ms SD, soft normalization, and 12 PCs. O'Doherty conditions combine 12 direction bins and three reach-magnitude bins; session-specific latent trajectories are aligned by orthogonal Procrustes and averaged. Neural windows cover execution for Monkey~1 (approximately 0--400\,ms from movement onset) and peri-movement activity for Monkeys~2--3 ($-200$ to $200$\,ms). These different windows reflect the task paradigms: Monkey~1 had an explicitly defined movement-execution phase in an instructed-delay task, whereas Monkeys~2--3 performed continuous, self-paced reaching without instructed epochs, motivating peri-movement windows centered on kinematically detected movement onset.

For centered cortical scores $Y_c$ and recorded hand-position and velocity variables $K_c$, we fit $B=\arg\min_B\|Y_cB-K_c\|_F^2$ and set $Z=Y_cU_BU_B^\top$, where $U_B$ spans the columns of $B$. The teacher has rank four, stored in 12 coordinates; it retains cortical variation within these directions, rather than replacing activity with kinematics. Construction uses all 36 conditions before student splits. Thus, eight conditions are held out only from student optimization. Because conditions include magnitude bins, withholding a condition need not remove its entire direction class. New evaluation directions lie outside the original condition set (Appendix~\ref{app:noveltargets}).

\paragraph{Control teachers.}
The task-only baseline and five alternative teachers probe different possible sources of benefit from alignment. Each teacher replaces the cortical representation in the same alignment objective; task-only omits this term.
\begin{itemize}
\item \textbf{Task-only:} the controller minimizes reaching loss alone ($\alpha=0$), without a teacher, using the same 28 training conditions and seed-specific split as the aligned controllers.
\item \textbf{Behavior:} this control supplies geometry derived directly from recorded movement rather than neural activity. We selected the \emph{Behavior} teacher as the strongest overall control among several candidates in preliminary tests, based on achieving the lowest error in the largest number of comparisons. It uses the same recorded $x,y,v_x,v_y$ variables that select the cortical subspace. We concatenate offsets of $-100,-50,0,50,100$\,ms, clip out-of-window indices to boundary samples, standardize each of the 20 features across all $36\times40$ teacher-fit condition--time observations, and retain four unwhitened PCs with deterministic signs. Positive offsets use future samples only in this offline teacher; recordings are unnecessary at inference. A fixed orthonormal map embeds the representation into 12 coordinates, preserving rank four without adding noise; it does not match the cortical covariance spectrum or effective alignment gradients. One frozen teacher per animal is shared across seeds, architectures, and doses. Monkey~1 uses the verified straight-reach execution trajectories underlying cortical-subspace selection. The behavioral representation preserves the original condition and sample ordering. For Monkeys~2--3, execution-window kinematics select a peri-movement neural subspace, so physical time windows are not matched.
\item \textbf{Ring:} this engineered teacher tests a simple directional prior, motivated by the directional organization observed in cortical activity and trained controllers. See Appendix~\ref{app:directional-ring} for the full analysis of directional organization, the control's motivation, and its construction and visualization. Its first two coordinates multiply the cosine and sine of the target angle by a shared temporal envelope, increasing linearly from 0.3 to 1.0 before Gaussian smoothing with a 30-ms SD\@. Each coordinate is scaled using the corresponding leading cortical covariance eigenvalue. Direction stays fixed within a reach while amplitude increases, so trajectories extend radially rather than rotate around the ring. Ten additional coordinates are generated from temporally smoothed Gaussian signals, centered, rotated to their principal axes, and scaled using the remaining cortical eigenvalues, independently for each model seed. Concatenating the directional and random components gives 12 coordinates; their cross-covariances prevent exact matching of the complete cortical covariance spectrum.
\item \textbf{Gaussian:} this control preserves population variance structure without imposing reach-direction geometry. Independent Gaussian signals are smoothed along time with a 30-ms SD, centered over all $36\times40$ samples, rotated to their principal axes, and rescaled using the cortical covariance eigenvalues. The resulting 12-coordinate teachers match the normalized cortical covariance spectrum and are generated independently per model seed. Spectrum matching applies to the pooled 36 conditions, not necessarily to each 28-condition training subset.
\item \textbf{Poisson:} this control supplies synthetic population activity using animal-specific fitted lognormal unit rates with Gaussian log-rate modulation (amplitude $0.4$, 30-ms smoothing). Twenty Poisson trials are averaged in 10-ms bins, retaining mean counts per bin rather than converting to rates in Hz. After 30-ms temporal smoothing, soft normalization by unit range plus five, centering, and unwhitened PCA over all 36 conditions, four PC scores are retained and eight exact-zero coordinates appended. One frozen teacher per animal and seed is shared across architectures and doses. This matches cortical teacher rank and stored dimension, not its covariance spectrum. Fitted rate summaries use execution windows, which differ from the peri-movement cortical windows for Monkeys~2--3.
\item \textbf{Shuffled:} this control disrupts condition correspondence while preserving complete cortical trajectories. An independent condition derangement is drawn for each animal and model seed, rejecting fixed points among all 36 conditions so that no trajectory remains assigned to its original condition. Complete trajectories are reassigned with their time order unchanged, and the permutation remains fixed throughout training and is shared across architectures and doses. Because different conditions can share a direction class, derangement of condition indices does not ensure that every assigned reach direction changes.
\end{itemize}

Table~\ref{tab:stochastic-controls} continues Table~\ref{tab:main} with Gaussian and Poisson columns, using the same metric, target sets, and training-based strength selection. Each control has complete 40-dose, 100-seed sweeps in both architectures and all three animals. For each control, 600 training-selected controllers were evaluated on the same 300 frozen new-target sets, yielding 4,800 new reaches. No retraining or strength retuning was performed.

\begin{table}[ht]
\caption{\textbf{Continuation of Table~\ref{tab:main}: Gaussian and Poisson controls.} Median $\pm$ bootstrap SE of final-window coordinate MSE (mm$^2$; 100 seeds, 100,000 seed resamples), using the same target sets and strength-selection protocol as Table~\ref{tab:main}. Each strength minimizes median training MSE over 40 doses and remains fixed across target sets; SEs condition on that selection. The sum of wins counts lowest medians across all seven methods in this table and Table~\ref{tab:main}, over the same 18 comparisons. Wins are descriptive, not significance tests.}
\label{tab:stochastic-controls}
\begin{center}
\begingroup
\setlength{\tabcolsep}{5pt}
\begin{tabular}{llrr}
\toprule
Student & Monkey & Gaussian & Poisson \\
\midrule
\multicolumn{4}{l}{\emph{A. Trained targets (28)}} \\
GRU & M1 & $54.85\pm3.46$ & $53.52\pm2.86$ \\
GRU & M2 & $52.97\pm2.79$ & $50.97\pm4.20$ \\
GRU & M3 & $56.55\pm3.64$ & $52.83\pm3.67$ \\
\addlinespace[2pt]
Transf. & M1 & $28.64\pm2.29$ & $26.70\pm3.59$ \\
Transf. & M2 & $18.52\pm2.17$ & $17.89\pm1.43$ \\
Transf. & M3 & $23.00\pm2.16$ & $18.73\pm2.22$ \\
\midrule
\multicolumn{4}{l}{\emph{B. Generalization to student-held-out targets (8)}} \\
GRU & M1 & $59.86\pm4.83$ & $60.23\pm4.35$ \\
GRU & M2 & $59.39\pm3.91$ & $52.54\pm2.57$ \\
GRU & M3 & $62.57\pm3.67$ & $59.59\pm4.12$ \\
\addlinespace[2pt]
Transf. & M1 & $35.12\pm4.33$ & $32.89\pm4.55$ \\
Transf. & M2 & $20.60\pm2.31$ & $19.61\pm1.78$ \\
Transf. & M3 & $25.43\pm3.44$ & $20.30\pm2.59$ \\
\midrule
\multicolumn{4}{l}{\emph{C. Generalization to new target directions (8)}} \\
GRU & M1 & $97.99\pm9.05$ & $88.68\pm6.23$ \\
GRU & M2 & $52.83\pm2.56$ & $49.32\pm3.49$ \\
GRU & M3 & $54.53\pm3.64$ & $49.53\pm2.19$ \\
\addlinespace[2pt]
Transf. & M1 & $119.49\pm14.44$ & $121.12\pm14.56$ \\
Transf. & M2 & $19.77\pm2.86$ & $19.88\pm1.65$ \\
Transf. & M3 & $22.16\pm2.94$ & $20.52\pm2.02$ \\
\midrule
\multicolumn{2}{l}{Sum of wins} & 0 & 0 \\
\bottomrule
\end{tabular}

\endgroup
\end{center}
\end{table}

\paragraph{Task and student architectures.}
Students control MotorNet's differentiable planar arm with two joints, six rigid-tendon Hill-type muscles, and 10-ms simulation steps \citep{codol2024motornet}. Rollouts last 40 steps and begin at the midpoint of joint ranges with zero joint velocity. Targets lie 10\,cm from the initial fingertip in each teacher's stored directions, irrespective of the recorded magnitude bin. Main-benchmark observations and rollouts are deterministic.

The closed-loop GRU \citep{cho2014rnn} has 128 hidden units and updated observations at every step. The open-loop Transformer \citep{vaswani2017attention} has three pre-normalized layers, width 64, four attention heads, feedforward width 128, and zero dropout. Initial observations, absolute and relative target coordinates, and target distance condition all time tokens, with learned positional embeddings and causal attention. It emits all 40 muscle commands in one forward pass without subsequent feedback; output biases initialize to $-2$. The Transformer is trained as an Inverter, following \citet{kapitonova2026inverse}, by backpropagating the reaching objective through the differentiable arm model over the full action sequence. The architectures are not parameter-count matched.

\paragraph{Population alignment objective.}
CCA rewards correlated linear projections across corresponding condition--time samples without neuron-to-unit assignments \citep{hotelling1936relations,andrew2013dcca}. We flatten training-condition and time axes, center each feature, and divide each activity matrix by its overall root-mean-square (RMS) magnitude. For normalized student states $H$ and teacher states $Z$, define $C_{HH}=H^\top H/(n-1)+\lambda I$, $C_{ZZ}=Z^\top Z/(n-1)+\lambda I$, and $C_{HZ}=H^\top Z/(n-1)$, with $\lambda=0.01$. The score is
\begin{equation}
 \overline{\rho}_\lambda(H,Z)=\frac{1}{12}\sum_{j=1}^{12}
 \operatorname{clip}_{[0,1]}\!\left[
 \sigma_j\!\left(C_{HH}^{-1/2}C_{HZ}C_{ZZ}^{-1/2}\right)\right].
 \label{eq:cca}
\end{equation}
Cholesky whitening and differentiable singular values implement the score; alignment gradients update only the student. Orthogonal coordinate changes preserve the score, whereas isotropic ridge regularization prevents invariance to arbitrary invertible transforms. Twelve singular values are averaged despite teacher rank four, giving an approximate ceiling of $4/12$. We retain this fixed normalization across teacher constructions rather than renormalizing by effective rank. For an exactly rank-four teacher, with its representation and regularization fixed, averaging all 12 rather than the leading four singular values divides the alignment term by three, a constant rescaling that can be absorbed into $\alpha$. Alignment rewards shared linear structure, not identical full-state distances or dynamical laws; functional benefit is evaluated through control performance \citep{kornblith2019similarity}.

\paragraph{Optimization and selection.}
For each seed $s=0,\ldots,99$, we randomly partition the 36 conditions into 28 training and eight evaluation conditions, using the same seed and split across teachers. All training conditions are optimized together for 700 Adam updates at learning rate $2\times10^{-3}$. Gradient norms are clipped to one; updates with nonfinite norms are skipped. For fingertip position $p_{ct}$, target $g_c$, and radius $R=0.10$\,m,
\begin{equation}
 \mathcal L_i=\frac{1}{2\cdot15\,|\mathcal C_{\rm tr}|R^2}
 \sum_{c\in\mathcal C_{\rm tr}}\sum_{t=26}^{40}
 \|p_{ct}-g_c\|_2^2
 -\alpha\min(1,i/60)\overline{\rho}_\lambda(H,Z).
 \label{eq:objective}
\end{equation}

\label{app:dosecoverage}
Neuro-Dist, Ring, Behavior, Gaussian, independently Shuffled, and Poisson teachers each have 40 positive strengths $0.1$--$4.0$ in steps of $0.1$, with 100 seeds per cell in both architectures and all three animals. Task-only uses $\alpha=0$. For Table~\ref{tab:main}, each teacher's strength minimizes median training coordinate MSE, with ties favoring the smaller strength. Models and strengths stay fixed across trained, student-held-out, and new targets. Gaussian and Poisson are additional controls outside the five-method main table; Supplementary Table~\ref{tab:stochastic-controls} uses the same strength-selection rule. The selected Shuffled models are shared by the noise and tangling analyses (Appendices~\ref{app:gruselectednoise} and~\ref{app:tangling}). All 100 seeds are retained in every cell, including high-error fits; no cells or runs are excluded.

Using 100 seeds per experimental setting allows us to characterize variability across model initializations and training/held-out condition splits. In the retrospective accuracy analysis, this also enabled separate groups of 50 seeds for alignment-strength selection and evaluation, with the evaluation seeds providing paired replication units for bootstrap confidence intervals and Holm-corrected comparisons (Appendix~\ref{app:accuracy-statistics}).

\paragraph{Evaluation and metrics.}
\label{app:metrics}
Clean-error tables and dose-response figures report final-window coordinate MSE,
\begin{equation}
 E_{\rm MSE}(\mathcal C)=\frac{10^6}{2\cdot15\,|\mathcal C|}
 \sum_{c\in\mathcal C}\sum_{t=26}^{40}\|p_{ct}-g_c\|_2^2
 \quad\text{(mm$^2$)}.
 \label{eq:framewise-mse}
\end{equation}
Here, $\mathcal C$ is the set of evaluated reaching conditions and $|\mathcal C|$ is its size.
This matches the reaching-loss reductions across the final 15 frames, conditions, and coordinates, in physical units and excluding alignment. Each model is scored separately on its 28 trained targets, eight student-held-out targets, or eight new directions; summaries are medians over all 100 seeds. New directions are constructed from target geometry alone, shared across methods, and at least 1 degree from every original direction up to numerical rounding (Appendix~\ref{app:noveltargets}). Sensory feedback noise evaluation uses coordinate RMSE, $\sqrt{E_{\rm MSE}}$, for each run and condition subset.

Tables~\ref{tab:main} and~\ref{tab:stochastic-controls} report the sample SD of $B=100{,}000$ bootstrap medians (denominator $B-1$). Each resample draws 100 whole-seed records with replacement, sharing the sampled seed indices across methods and target sets within each animal--controller group. Strengths remain fixed at their training-selected values. These SEs quantify uncertainty in the median across seeds, conditional on the recorded datasets and selected strengths; they exclude dose-selection uncertainty and are not paired tests of between-method differences.

The illustrative reaching examples report settled-endpoint error,
\begin{equation}
 E(\mathcal C)=1000\sqrt{\frac{1}{|\mathcal C|}
 \sum_{c\in\mathcal C}\left\|\frac{1}{15}\sum_{t=26}^{40}p_{ct}-g_c\right\|_2^2}
 \quad\text{(mm)}.
 \label{eq:reach}
\end{equation}
Averaging positions before measuring error omits terminal temporal variability; this endpoint metric also lacks the coordinate factor used by MSE and RMSE\@. Seeds vary initialization and condition split, not animal or recording session. Interquartile-range bands describe run dispersion, not uncertainty; differences of marginal medians are not medians of paired differences. Figure~\ref{fig:dose} scores trained targets and Figure~\ref{fig:dosegeneralization1} student-held-out targets. Their stars mark table-row winners at training-selected doses, not significance or held-out curve minima. Logarithmic axes and clipping affect display only; outcomes from all 100 seeds enter summaries.

\paragraph{Trajectory tangling.}
Following \citet{russo2018untangled}, nearby states with dissimilar velocities have high tangling. For condition--time states $x_i$, finite-difference velocities $\dot{x}_i$, and total variance $s_x^2=N^{-1}\sum_i\|x_i-\bar{x}\|_2^2$,
\begin{equation}
 Q_i=\max_j\frac{\|\dot{x}_i-\dot{x}_j\|_2^2}{\|x_i-x_j\|_2^2+0.1s_x^2},
 \qquad Q_S=\frac{1}{|S|}\sum_{i\in S}Q_i.
 \label{eq:tangling}
\end{equation}
The native 128-dimensional GRU states are used without PCA\@. This measures the full recurrent-state geometry without choosing a variance-retention threshold: PCA truncation can alter state and velocity distances and discard low-variance directions relevant to tangling. Training and held-out query subsets $S$ share all 36 conditions as references and for variance normalization. Restricting partners to the same or other reaches defines within- and between-reach tangling; total pointwise $Q$ is their maximum. Appendix~\ref{app:tangling} specifies derivatives, uncertainty, cortical references, new-target reference pools, and velocity/distance and speed/angular decompositions.

\paragraph{Sensory feedback noise evaluation.}
The main evaluation uses 2,100 frozen GRUs: seven methods, three animals, and 100 seeds, selected using clean training MSE\@. Independent Gaussian noise perturbs all 16 observations, including target information, at every 10-ms step. Four positive amplitudes $\sigma\in\{0.025,0.05,0.075,0.1\}$ each use three matched realizations. Coordinate RMSE scores only the eight student-held-out targets. Appendix~\ref{app:gruselectednoise} gives the dose selection, sampling, paired ratio endpoint, multiple-comparison correction, and uncertainty protocol.

\section{Supplementary experimental protocols}
\label{app:protocols}

\subsection{Student-held-out dose responses}
\begin{figure}[t]
\begin{center}
\includegraphics[width=\linewidth]{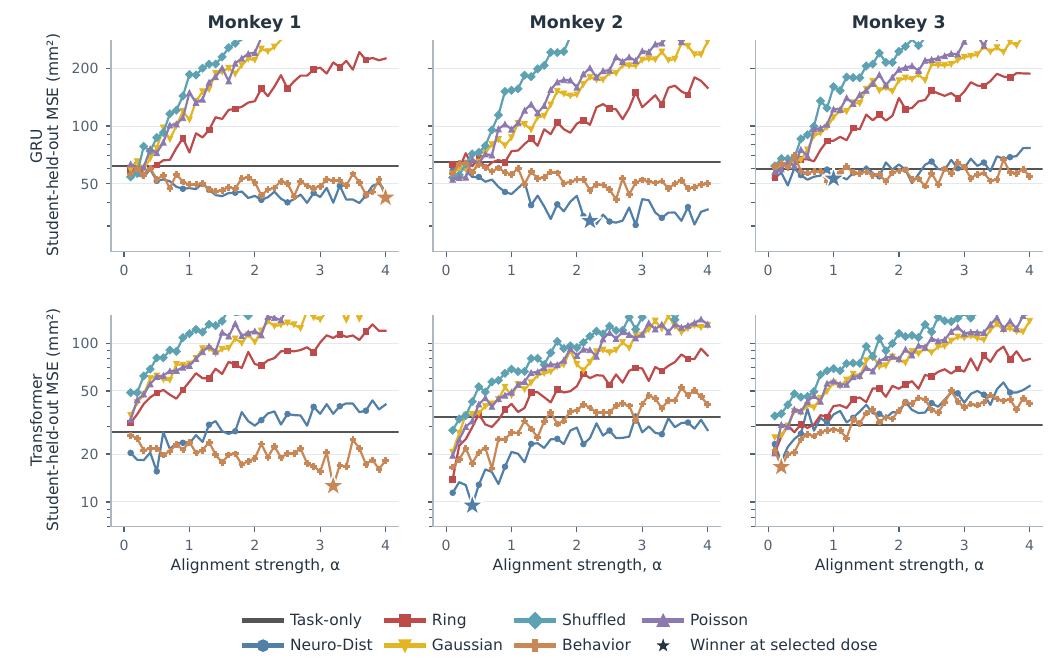}
\end{center}
\caption{\textbf{Neuro-Dist gains extend to student-held-out targets.} Median coordinate MSE on eight targets excluded from student training but included in teacher construction; 100 seeds per dose, without uncertainty bands. Stars compare the five methods in Table~\ref{tab:main} at training-selected strengths. Gaussian and Poisson are additional full-grid controls. The zoom clips off-scale values.}
\label{fig:dosegeneralization1}
\end{figure}

Figure~\ref{fig:dosegeneralization1} shows the full dose sweeps on the eight original conditions withheld from student optimization, with stars at the training-selected strengths used for Table~\ref{tab:main}. These conditions participated in teacher construction. Every aligned teacher has 40 positive strengths and all 100 seeds are retained at each strength, with no exclusions. Stars compare the five table methods at their selected doses, rather than minima chosen on held-out error. Gaussian and Poisson are additional plotted controls. The logarithmic display clips off-scale values; full numerical values are retained in the source artifacts.

\subsection{New-target dose responses}
\begin{figure}[htbp]
\begin{center}
\includegraphics[width=\linewidth]{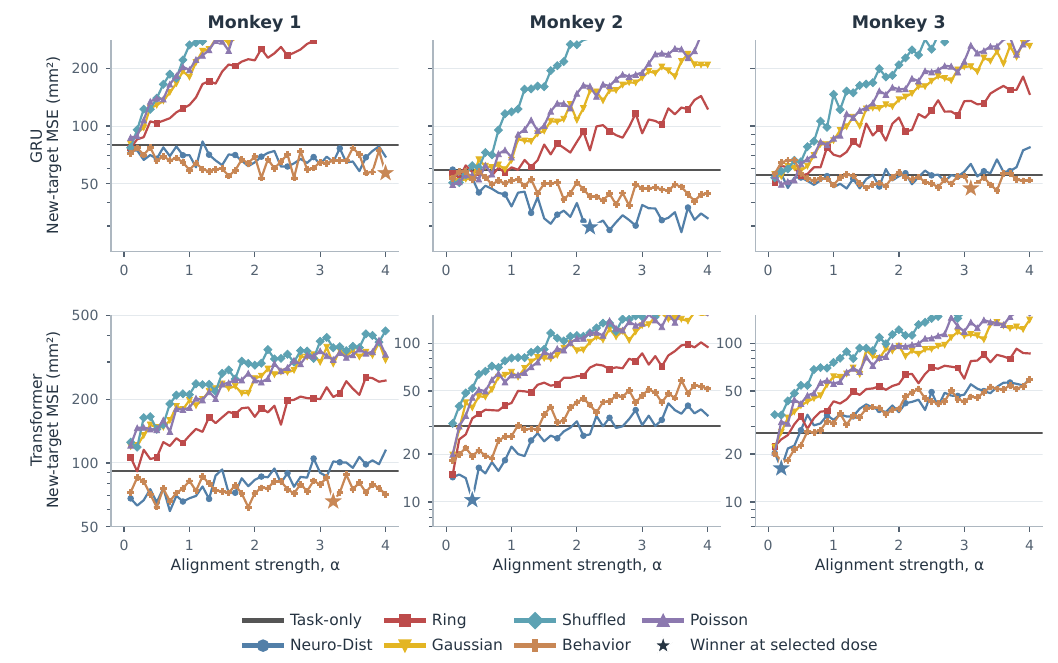}
\end{center}
\caption{\textbf{Dose responses on new target directions.} Median coordinate MSE on eight directions absent from both student training and teacher construction; 100 seeds per dose, without uncertainty bands. Stars compare the five methods in Table~\ref{tab:main} at their unchanged training-selected strengths. Gaussian and Poisson are additional full-grid controls.}
\label{fig:dosegeneralization2}
\end{figure}

Figure~\ref{fig:dosegeneralization2} evaluates all 144,600 saved GRU and Transformer controllers on the eight new directions defined in Appendix~\ref{app:noveltargets}. Each of the six aligned teachers has 40 strengths and 100 seeds per strength, alongside the shared task-only baseline. These descriptive curves retain the original training-selected operating points; no new-target outcome selects a strength or changes the statistical comparisons.

\subsection{Generalization to new target directions}
\label{app:noveltargets}
We construct eight additional directions for each animal and seed, yielding 300 target sets shared across methods and architectures. Construction uses only original target directions and condition splits, not neural activity or controller outcomes. With the eight original held-out directions fixed, we maximize minimum circular separation over old--new and new--new pairs by enumerating the 6,435 allocations of eight new targets among eight cyclic gaps. Fixed old--old pairs do not enter the objective. New directions then receive the smallest displacement needed to remain at least 1 degree from every original direction, with deterministic tie-breaking and a $10^{-4}$-degree numerical tolerance. The arm retains the same start posture and 10-cm target radius.

The 3,000 controllers and their strengths in Table~\ref{tab:main} are frozen before evaluation, producing 24,000 new reaches. Each new rollout resets controller and environment state and includes the eight new targets alongside the original 28 training targets; only the new targets contribute to the reported error. No teacher refitting, retraining, dose retuning, or model exclusions occur.

These are directions outside the discrete teacher-construction and student-training sets, sometimes only 1 degree from an original direction. This evaluates within-task target transfer, not a teacher's predictions or transfer across sectors, sessions, tasks, or postures. Figure~\ref{fig:dosegeneralization2} extends the same evaluation protocol to the full dose grid; the selected-strength comparisons remain unchanged.

\subsection{Overview figure and example selection}
\label{app:overview}
Figure~\ref{fig:framework}B uses Monkey~2 session \texttt{indy\_20160627\_01}, a contributing session rather than the 37-session pooled training teacher. The cortical trace display uses unsmoothed, condition-averaged 10-ms spike counts from 219 nonempty sorted primary motor cortex units, without additional isolation-quality filtering. Ten displayed units are individually normalized by their maxima. The population paths instead retain the teacher's 30-ms Gaussian smoothing and this session's frozen projection. Twelve middle-magnitude direction conditions are displayed; three PCs explain 90.5\% of rank-four subspace variance across all 36 conditions, not all firing-rate variance.

The paired population example is a Neuro-Dist GRU (seed~71, $\alpha=2.2$). Ten traces use one common $2\times$ display gain and vertical offsets; all 128 units enter the independent PCA, whose first three PCs explain 63.6\% of hidden-state variance. Cortical bins span $-200$ to $200$\,ms relative to movement onset; GRU post-update states span 10--400\,ms and are displayed relative to the rollout midpoint. Matching display times or colors therefore do not imply physical-time correspondence. The connector illustrates the CCA training objective; the independent display projections are not a fitted CCA alignment.

Panel~C shows examples selected to represent the observed reaching behavior: Monkey~2 GRU seed~53 ($\alpha=3.6$) and Transformer seed~61 ($\alpha=0.4$). Strengths were selected by minimizing population median training settled-endpoint error. All eight student-held-out reaches are shown at a common spatial scale; the two seeds have different splits. Filled dots denote final-window mean positions, and printed errors use Equation~\ref{eq:reach}, rather than framewise MSE\@. 

\section{GRU sensory feedback noise on student-held-out targets}
\label{app:gruselectednoise}

\begin{figure}[htbp]
\begin{center}
\includegraphics[width=\linewidth]{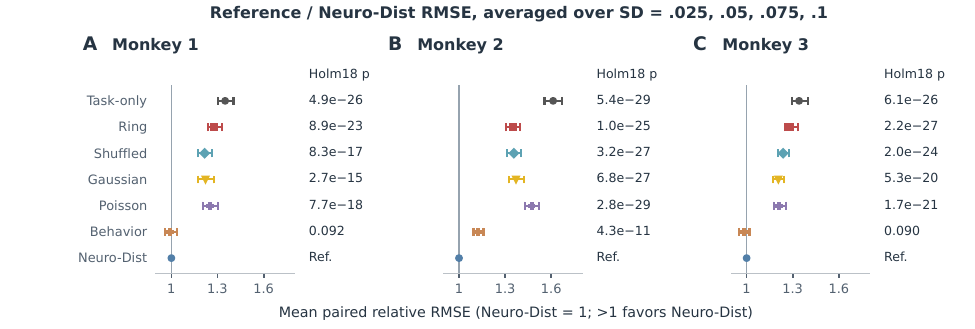}
\end{center}
\caption{\textbf{Primary comparisons of noisy student-held-out accuracy.} (A--C) Monkeys~1--3 for the same 2,100 selected-strength GRUs as in Figure~\ref{fig:sensorynoise}. Strengths minimize clean error on 28 student-training targets; evaluation scores the eight student-held-out targets, included in all-36-condition teacher construction. Three noise realizations are averaged per amplitude. Reference errors are divided by the Neuro-Dist error for the same seed separately at each positive amplitude ($\sigma=0.025,0.05,0.075,0.1$), then averaged within seed. Markers show the mean ratio across 100 seeds; values above one favor Neuro-Dist. Whiskers are pointwise 95\% BCa intervals from 50,000 seed resamples. Neuro-Dist's self-reference is exactly one and has no independent interval or test. Primary paired signed-rank $p$-values use one Holm correction across all six references and three animals (18 tests).}
\label{fig:grunoiseheldoutcomparisons}
\end{figure}

\paragraph{Cohort and strength selection.}
Figure~\ref{fig:sensorynoise} and Supplementary Figure~\ref{fig:grunoiseheldoutcomparisons} evaluate 2,100 frozen GRUs: seven methods, three animals, and 100 seeds. For each teacher and animal, one strength minimizes median clean final-15-frame coordinate MSE on the 28 training conditions, with ties favoring the smaller strength. No noisy, held-out, or per-seed outcome selects the dose. All candidate cells contain 100 seeds, and all 2,100 selected models have finite clean and noisy errors; no cells or models are excluded. Neuro-Dist, Ring, Gaussian, Behavior, Poisson, and Shuffled have 40 strengths ($0.1$--$4.0$, step $0.1$).

For Monkeys~1/2/3, selected strengths are $1.5/2.2/1.0$ for Neuro-Dist, $0.1/0.6/0.1$ for Ring, $0.2/0.2/0.1$ for Shuffled, $0.3/0.3/0.2$ for Gaussian, $0.2/0.1/0.2$ for Poisson, and $4.0/2.6/3.1$ for Behavior; task-only uses zero. The original seed permutation defines the 28/8 split, matched across teachers. All 36 conditions enter teacher construction. No models were retrained or doses reselected for this decomposition.

\paragraph{Perturbation and scoring.}
Independent Gaussian noise perturbs all 16 observation channels at every 10-ms feedback step: two target coordinates, two visual fingertip coordinates, and 12 muscle-length and muscle-velocity channels, normalized by optimal fiber length and maximum shortening velocity, respectively. The four positive amplitudes are $\sigma\in\{0.025,0.05,0.075,0.1\}$, with three realizations per model seed, matched across methods and amplitudes. Positional coordinates are in meters, so their SDs are 2.5--10\,cm for a 10-cm reach; muscle channels have different normalized units. Noise is not demeaned or rescaled by realized RMS\@.

The rollout includes all 36 original targets and lasts 40 simulation steps. Only the eight student-held-out targets contribute to the final-15-frame coordinate RMSE, averaged across realizations. Clean error is evaluated separately. Figure~\ref{fig:sensorynoise}A is a separate fixed-dose, 1.5-s illustration; quantitative comparisons use these 400-ms selected-strength rollouts.

\newpage
\paragraph{Paired inference.}
For each reference $j$, animal $a$, and model seed $s$, we compute
\[
q_{asj}=\frac14\sum_{\sigma\in\{0.025,0.05,0.075,0.1\}}
\frac{E_{asj}(\sigma)}{E_{as,\mathrm{Neuro\text{-}Dist}}(\sigma)},
\]
where each $E$ first averages realization-level RMSE\@. Thus, normalization precedes averaging and positive levels receive equal weight. Values above one favor Neuro-Dist; this score measures relative noisy accuracy rather than degradation from clean performance.

Eighteen two-sided signed-rank tests (six references by three animals) share one Holm correction. Pointwise 95\% bias-corrected and accelerated (BCa) intervals for mean ratios use 50,000 paired-seed resamples, preserving all amplitudes within each seed. The statistical unit is the model seed ($n=100$), not a target, frame, amplitude, or realization. Signed-rank inference assumes symmetry and does not directly test mean ratios.

\paragraph{Results and scope.}
The signed-rank tests favor Neuro-Dist over task-only, Ring, Shuffled, Gaussian, and Poisson in every animal; all 15 comparisons have $p\leq2.68\times10^{-15}$ after Holm18 correction including Behavior. Mean ratios of task-only error to Neuro-Dist error are 1.350/1.612/1.341 in Monkeys~1/2/3. Neuro-Dist has lower mean error than task-only at every positive amplitude; averaging amplitudes gives 30.65/26.45/31.49\,mm versus 40.74/42.28/41.88\,mm.

Behavior has mean ratios 0.991/1.125/0.983. The signed-rank tests favor Neuro-Dist in Monkey~2 (corrected $p=4.33\times10^{-11}$) and detect no statistically significant difference in Monkeys~1 and~3 ($p=0.0924$ and $0.0897$, respectively). Behavior mean-ratio intervals include one in Monkeys~1 and~3: $[0.9578,1.0360]$ and $[0.9522,1.0185]$.

Neuro-Dist also has smaller mean absolute increases from clean error than task-only in all 12 animal--amplitude combinations. Mean increases across amplitudes are 23.92/20.72/23.94\,mm versus 32.26/33.16/33.58\,mm. These increases are descriptive; the paired tests concern noisy accuracy, not uniformly smaller relative degradation. Comparison with Behavior remains dataset dependent. Inference conditions on frozen strengths: the same seeds enter selection and estimation, so selection uncertainty is not propagated.

\section{Directional organization and the engineered ring control}
\label{app:directional-ring}

\subsection{Directional organization motivates a simple control}
\begin{figure}[!htb]
\centering
\includegraphics[width=\linewidth]{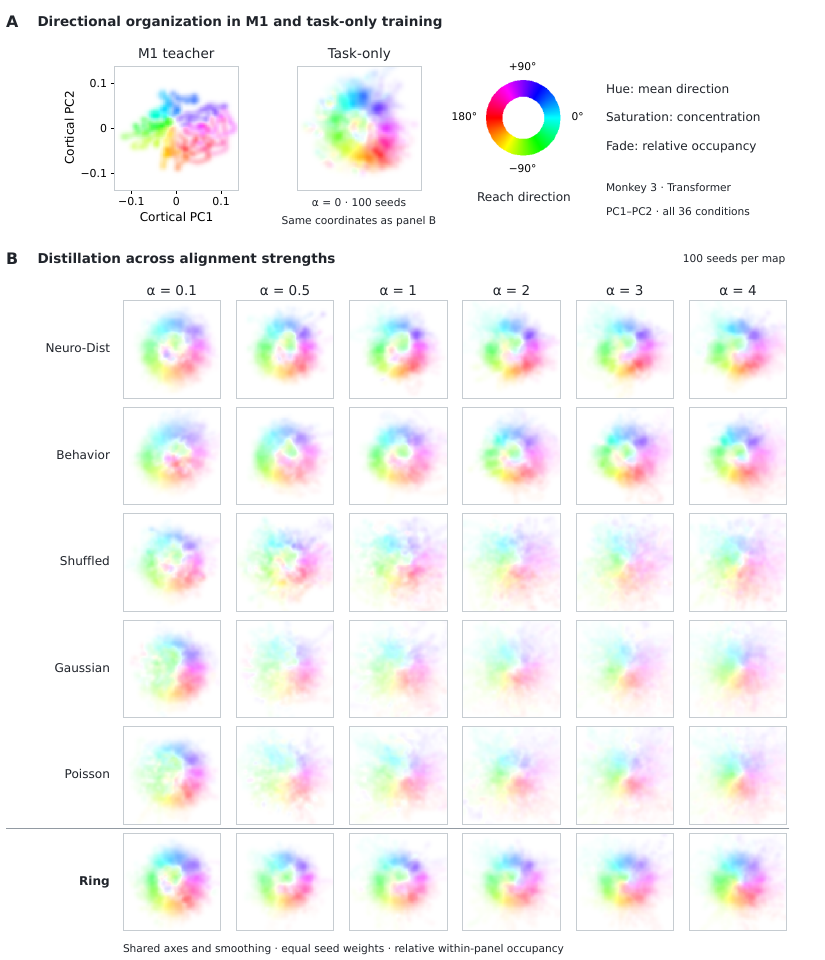}
\caption{\textbf{Directional organization motivating the engineered Ring control.} In this figure only, M1 denotes primary motor cortex. \textbf{A:} Monkey~3 M1 teaching representation and task-only Transformer states in common cortical PC1--PC2 coordinates. \textbf{B:} Transformer states across six illustrative alignment strengths. Every student map pools 100 seeds and all 36 conditions; alignment is fitted on the 28 training conditions per seed. Hue indicates reach direction, saturation directional concentration, and fading relative occupancy within each panel. The same cortical reference, axes, and smoothing apply throughout. Viewport cropping and projection details are specified in Appendix~\ref{app:directional-ring}. Directional organization is visible under multiple training regimes and is not specific to cortical supervision.}
\label{fig:directional-organization}
\end{figure}

Figure~\ref{fig:directional-organization} illustrates reach-direction organization in the Monkey~3 primary motor cortex teaching representation and Transformer states. Neighboring regions in the displayed state space preferentially represent different reach directions. This organization is visible in the cortical reference, task-only controllers, and controllers trained with Neuro-Dist or Behavior supervision. It therefore motivates a control that supplies a simple, explicitly constructed directional geometry. The Ring teacher asks whether this accessible feature of task-related population organization is sufficient to account for benefits of cortical supervision; its functional consequences are evaluated by the accuracy and robustness experiments, rather than inferred from the appearance of these maps.

The example also shows directionally organized Ring-trained states. Thus, directional organization is shared across several training regimes and is not a signature unique to Neuro-Dist. The displayed cortical reference is the behavior-selected rank-four primary motor cortex representation used for teaching, not the complete recorded population.

\subsection{Construction of the directional density maps}
For each animal we fit an unwhitened four-dimensional PCA basis to the fixed cortical teacher. Within each model seed, student PCA and a four-dimensional orthogonal Procrustes alignment are fitted using only the 28 training conditions. A single scalar matches the centered four-dimensional training-state RMS magnitude to the cortical reference; translation matches the cortical training mean. The fitted transformation is then applied to all 36 conditions, without per-axis scaling or nonlinear warping. Figure~\ref{fig:directional-organization} displays the first two cortical coordinates, pooling all 40 frames of every condition. Each student panel includes all 100 seeds with equal weight.

For occupancy $D$ and complex directional mass $R$, each condition--time location contributes equal probability weight $1/(36\times40)$ to $D$ and this weight times $\exp(\mathrm{i}\theta_c)$ to $R$. We average these arrays across seeds and smooth both arrays on a $128\times128$ grid with a Gaussian kernel of standard deviation 2 pixels. Hue is the circular mean direction $\arg R$, saturation is the local concentration $|R|/D$, and color fades toward white according to relative occupancy $(D/\max D)^{0.55}$. The maximum is taken separately within each panel, so brightness encodes occupancy relative to that panel's peak.

All displayed maps use one common square viewport derived from the pooled 99.5th-percentile maximum absolute coordinate, $\max(|\mathrm{PC1}|,|\mathrm{PC2}|)$, across both architectures, all teachers and strengths, enlarged by 10\% and expanded to include every cortical reference point. In this figure, at most 0.496\% of pooled state samples lie outside any one panel; the largest individual-seed fraction outside is 10.14\%. No seed is excluded and cropped probability is not renormalized. The cortical teacher was constructed from all 36 conditions, including those withheld from student training.

\begin{figure}[!htb]
\centering
\includegraphics[width=\linewidth]{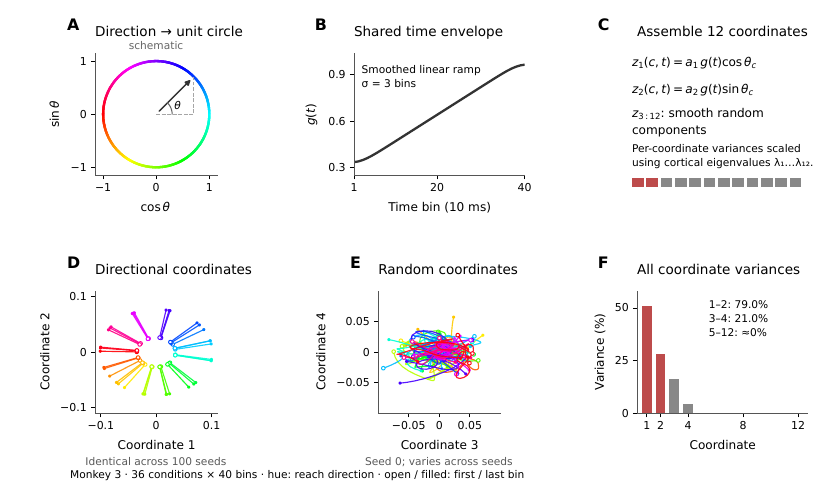}
\caption{\textbf{The engineered Ring teacher: schematic and numerical construction.} \textbf{A:} Idealized direction-to-circle mapping. \textbf{B:} Actual shared temporal envelope. \textbf{C:} Twelve-coordinate construction. \textbf{D:} Actual directional coordinates for all 36 Monkey~3 conditions, identical across seeds. \textbf{E:} Smooth random coordinates 3--4 for seed~0. \textbf{F:} Variance of all 12 coordinates. Hue denotes reach direction as in Figure~\ref{fig:directional-organization}; open and filled markers identify the first and last frames. Panels D--F show the numerical teacher from the training implementation. Its first pair is anisotropically scaled and evolves radially, while the remaining components provide smooth random structure. These are construction coordinates, not PCs of the full teacher.}
\label{fig:directional-ring-teacher}
\end{figure}

\subsection{From an idealized circle to the actual engineered teacher}
Figure~\ref{fig:directional-ring-teacher} separates the schematic directional construction from an actual teacher generated by the training implementation. For condition angle $\theta_c$ and frame $t$, the directional pair is
\begin{equation}
 z_{ct,1}=a_1g_t\cos\theta_c,\qquad
 z_{ct,2}=a_2g_t\sin\theta_c,
 \label{eq:ring-direction-pair}
\end{equation}
where $g$ is a linear envelope from 0.3 to 1.0, smoothed with a Gaussian kernel of standard deviation 3 frames. Each $a_d$ scales the variance of its directional coordinate using the corresponding leading cortical covariance eigenvalue. The angle is constant within each reach while the amplitude increases; trajectories extend radially rather than rotate around the ring. Because the two scale factors differ, the direction-indexed curve is generally an ellipse in construction coordinates.

Ten additional coordinates are obtained from temporally smoothed Gaussian signals, centered, rotated to their principal axes, and scaled using the remaining cortical eigenvalues. These components are generated independently for each model seed and concatenated with the directional pair to give the 12-coordinate Ring teacher. Cross-covariances with the directional coordinates mean that this construction does not exactly match the complete cortical covariance spectrum. In the displayed Monkey~3 seed-0 teacher, the first pair accounts for approximately 79.0\% of total coordinate variance and coordinates 3--4 for 21.0\%; coordinates 5--12 contain only numerical residual variance. The plotted axes are construction coordinates, not PCs of the complete concatenated teacher.

The displayed teacher uses the first seed, without performance-based selection, and is generated from the cortical reference using the same construction as in training. The schematic circle explains the design principle; the numerical panels show the actual scaled directional paths and stochastic remainder. This construction provides a transparent test of a simple directional prior, without assuming that cortical population activity follows the engineered trajectories or their temporal envelope.

\section{Trajectory-tangling analysis}
\label{app:tangling}

We analyzed all 2,100 GRUs spanning seven methods, three animals, and 100 seeds per cell (Figure~\ref{fig:gru-tangling}). Aligned strengths minimize median training MSE across the 40-dose grid. Neuro-Dist reduces median training tangling by 48--68\% relative to task-only. The Neuro-Dist--Behavior comparison favors Neuro-Dist in Monkeys~1--2 and Behavior in Monkey~3 across trained, student-held-out, and new targets.

\begin{figure}[t]
\begin{center}
\includegraphics[width=0.9\linewidth]{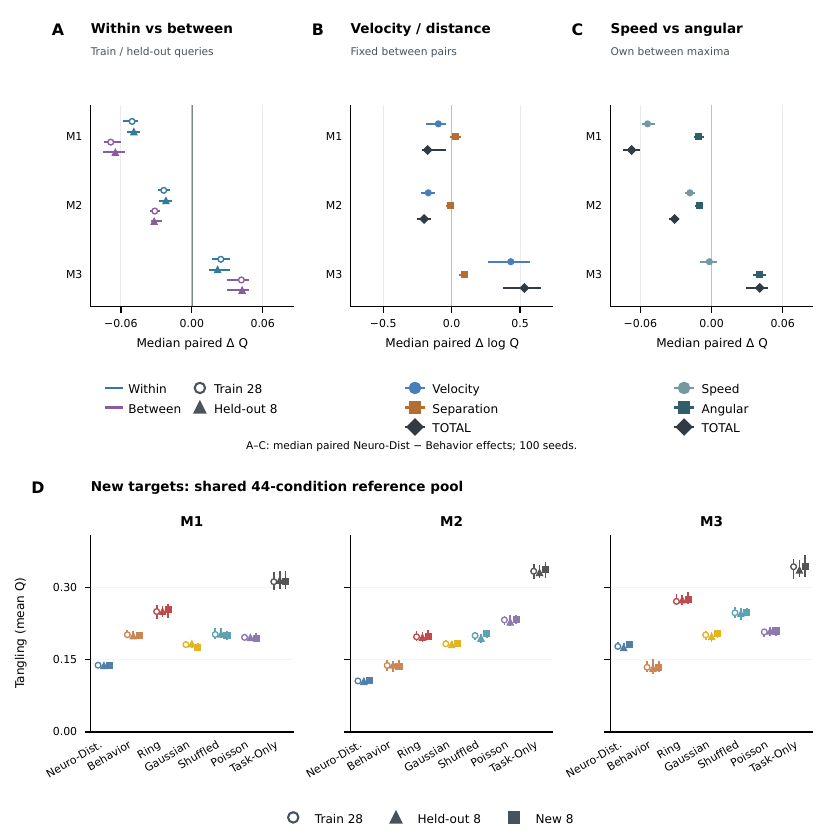}
\end{center}
\caption{\textbf{Tangling differences persist at new target directions.} (A--C) Median seed-paired Neuro-Dist-minus-Behavior differences: within- and between-reach $Q$ (A), velocity and separation contributions at shared task-only-selected pairs (B), and speed and angular contributions at each model's maximizing pairs (C). Negative separation effects indicate larger normalized denominators; component medians need not sum to the total. A uses trained and held-out queries; B--C use all 36. (D) Medians across seeds of model-level mean total $Q$ for all seven methods at trained, held-out, and new targets, using a common 44-condition reference pool and variance instead of the original 36. Whiskers are pointwise 95\% bootstrap intervals for the medians across 100 seeds, conditional on selected doses and fixed datasets.}
\label{fig:gru-tangling-supplement}
\end{figure}

We compute Equation~\ref{eq:tangling} in all 128 hidden coordinates, using centered interior and one-sided endpoint differences with $\Delta t=1$ frame. All $36\times40$ original states define the reference pool and variance normalization. Within-model means summarize total, within-reach, and between-reach tangling separately for trained and held-out queries; pointwise total $Q$ is the maximum of the latter two components. We separate velocity disagreement from state separation at shared condition--time pairs selected using task-only models, and resolve velocity disagreement into speed and angular contributions at each model's maximizing pairs (Figure~\ref{fig:gru-tangling-supplement}A--C).

For new-target analysis, the same controllers reach eight additional directions shared across methods for each animal and seed. We recompute $Q$ using a common pool of 44 original and new conditions and its variance, then summarize trained, held-out, and new queries separately (Figure~\ref{fig:gru-tangling-supplement}D). A sensitivity analysis compares held-out and new queries against the 28 training conditions with fixed original-condition variance.

We report medians across seeds and medians of paired seed-wise differences. Pointwise 95\% percentile intervals use 10,000 bootstrap resamples shared across methods, preserving model--target pairing and conditioning on selected strengths and fixed datasets. Across the 21 animal--method summaries, we assess the descriptive tangling--error association with Spearman correlation and show an ordinary least-squares fit.

Cortical $Q$ provides a descriptive reference from the parent 12-PC representation, with session pooling for Monkeys~2--3 and distinct temporal preprocessing. Figure~\ref{fig:tangling} illustrates a representative Monkey~2 session and matched GRU pair (seed~45) using independent PCA projections of 12 representative reaches; its printed $Q$ uses all 36 conditions in the native representations.

\section{Statistical comparison of reaching accuracy}
\label{app:accuracy-statistics}

\subsection{Methods}

\paragraph{Question and analysis scope.}
We compare Neuro-Dist with each of Task-only, Ring, Behavior (delayed position--velocity), Shuffled, Gaussian, and Poisson in GRUs and Transformers across the three animal datasets. The outcome is each controller's final-window coordinate MSE (Equation~\ref{eq:framewise-mse}): squared position residuals averaged over the final 15 post-action frames, both coordinates, and evaluated targets, in mm$^2$. All evaluated MSEs are finite, and all runs are retained, including high-error fits. Tables~\ref{tab:main} and~\ref{tab:stochastic-controls} report descriptive medians and bootstrap SEs across all 100 seeds at training-selected strengths. These SEs condition on the selected strengths and do not correct for reuse of the same seeds for selection and estimation \citep{cawley2010overfitting}; they supply no selection-adjusted significance claims. For statistical comparisons, we use a \emph{seed-split validation}, separating strength selection from evaluation.

\paragraph{Independent strength-selection seeds.}
Before computing validation contrasts, we fixed a deterministic partition of the 100 seeds into 50 selection and 50 evaluation seeds, shared across methods, animals, and architectures. For each teacher, animal, and architecture, the selected $\alpha$ minimizes median MSE on the 28 trained targets among 40 strengths $0.1{:}0.1{:}4.0$, using only selection seeds and resolving ties toward smaller strengths. Task-only retains the original 28-target baseline at $\alpha=0$. The resulting dose map differs from the table's map in 16 of 42 cells (Table~\ref{tab:accuracy-statistics-doses}). Each strength stays fixed across evaluation seeds, target regimes, and bootstrap draws; it is not optimized per seed or for held-out/new-target error.

\begin{table}[htbp]
\caption{\textbf{Frozen strengths for seed-split validation.} Each $\alpha$ minimizes trained-target median MSE on the 50 selection seeds; results use the other 50. Parentheses give the original 100-seed table strength where it differs. Each selected strength applies unchanged to all three target regimes. M1--M3 denote Monkeys~1--3.}
\label{tab:accuracy-statistics-doses}
\begin{center}
\begingroup
\setlength{\tabcolsep}{3pt}
\begin{tabular}{@{}lccccccc@{}}
\toprule
Cohort & \shortstack{Task-\\only} & Ring & \shortstack{Neuro-\\Dist} & Behavior & \shortstack{Shuf-\\fled} & \shortstack{Gaus-\\sian} & Poisson \\
\midrule
GRU/M1 & 0.0 & 0.3 (0.1) & 2.4 (1.5) & 2.3 (4.0) & 0.2 & 0.3 & 0.2 \\
GRU/M2 & 0.0 & 0.6 & 2.2 & 1.8 (2.6) & 0.2 & 0.4 (0.3) & 0.1 \\
GRU/M3 & 0.0 & 0.1 & 1.4 (1.0) & 2.8 (3.1) & 0.3 (0.1) & 0.1 (0.2) & 0.2 \\
Transf./M1 & 0.0 & 0.2 (0.1) & 0.3 & 3.7 (3.2) & 0.1 & 0.1 & 0.1 \\
Transf./M2 & 0.0 & 0.1 & 0.1 (0.4) & 0.7 (0.1) & 0.2 (0.1) & 0.1 & 0.1 \\
Transf./M3 & 0.0 & 0.2 (0.1) & 0.2 & 0.2 & 0.1 (0.2) & 0.1 & 0.1 \\
\bottomrule
\end{tabular}

\endgroup
\end{center}
\end{table}

\paragraph{Three evaluation regimes.}
The other 50 seeds supply results on the 28 trained targets, eight targets withheld from student training but included in teacher construction, and eight directions unseen by both teacher and student. The last regime uses the fixed animal/seed target bundles of Appendix~\ref{app:noveltargets}. The 2,100 evaluation-half controllers and their selected strengths remain fixed across all three regimes, with no retraining, teacher refitting, or exclusions. The new directions remain within the original task, radius, and start posture.

\paragraph{Factorial median model and effects.}
We estimate a saturated raw-scale median model with one parameter for every method $j$, architecture $a$, animal $m$, and target regime $r$: $7\times2\times3\times3=126$ cells. This is a fully interacted median-regression model \citep{koenker2001quantile}, fitted directly by cell medians, without imposing homogeneous teacher effects. For even sample sizes, the median is the arithmetic mean of the two central observations. With $\widehat\mu_{jamr}$ denoting a raw MSE median, we define
\begin{equation}
 d_{jamr}=\log\!\frac{\widehat\mu_{\mathrm{ND},amr}}{\widehat\mu_{jamr}},
 \qquad
 D_{jr}=\frac{1}{6}\sum_{a,m}d_{jamr},
 \qquad
 G_{jr}=100\bigl(1-e^{D_{jr}}\bigr).
 \label{eq:accuracy-statistics-effect}
\end{equation}
Positive $G$ favors Neuro-Dist. This is an equally weighted geometric-average median-MSE reduction across the six animal--architecture cells. We take logs \emph{after} raw-scale medians. For percentages referenced to Neuro-Dist, the equivalent contrast is $H_{jr}=100(e^{-D_{jr}}-1)$: positive values indicate increased control MSE, and zero indicates equal performance. The combined dose/comparison figure uses this scale, transforming the log-interval endpoints in reverse order; the tables retain $G_{jr}$.

\begin{table}[htbp]
\caption{\textbf{Aggregate seed-split validation.} Geometric-average median-MSE reduction (\%) and pointwise 95\% BCa interval versus each comparator. All three target regimes form one 18-comparison family; $\dagger$ marks Holm-adjusted $p<0.05$, not simultaneous interval coverage. Results use 50 evaluation seeds with independently selected, frozen strengths.}
\label{tab:accuracy-statistics-aggregate}
\begin{center}
\begingroup
\setlength{\tabcolsep}{3pt}
\begin{tabular}{@{}lccc@{}}
\toprule
Comparator & Trained targets & Student-held-out & New targets \\
\midrule
Task-only & $42.2\;[29.3,\,53.0]^{\dagger}$ & $30.7\;[8.9,\,42.9]^{\dagger}$ & $27.5\;[8.6,\,38.4]^{\dagger}$ \\
Ring & $36.8\;[24.5,\,48.0]^{\dagger}$ & $29.5\;[12.5,\,40.0]^{\dagger}$ & $26.8\;[14.5,\,37.8]^{\dagger}$ \\
Behavior & $26.9\;[14.7,\,41.0]^{\dagger}$ & $15.2\;[-2.6,\,29.7]$ & $16.7\;[5.0,\,28.3]^{\dagger}$ \\
Shuffled & $51.1\;[42.9,\,59.7]^{\dagger}$ & $43.5\;[29.5,\,52.0]^{\dagger}$ & $44.5\;[36.4,\,50.8]^{\dagger}$ \\
Gaussian & $41.8\;[31.1,\,53.4]^{\dagger}$ & $34.6\;[20.0,\,46.8]^{\dagger}$ & $28.9\;[14.0,\,35.7]^{\dagger}$ \\
Poisson & $38.9\;[30.6,\,50.3]^{\dagger}$ & $29.1\;[12.9,\,40.3]^{\dagger}$ & $23.2\;[10.0,\,32.1]^{\dagger}$ \\
\bottomrule
\end{tabular}

\endgroup
\end{center}
\end{table}

\paragraph{Paired bootstrap and multiplicity.}
One whole-seed index vector is resampled with replacement and shared across every method, animal, architecture, and target regime. This preserves dependencies from initialization, condition splits, and seed-specific new-target sets; neither frames nor reaches are independent replications. The validation uses 4,000,000 resamples of 50 evaluation blocks; the selection half remains fixed. We compute pointwise 95\% bias-corrected and accelerated (BCa) intervals \citep{efron1987better}, using tie-aware bias correction (the average strict/weak bootstrap rank), paired leave-one-seed-out acceleration, and linear quantile interpolation. Two-sided marginal $p$-values invert these confidence curves: the smallest $\alpha$ for which the $1-\alpha$ interval excludes zero, preserving strict exclusion at ties. Values below $10^{-5}$ are conservatively reported as upper bounds. Holm's step-down correction \citep{holm1979simple} uses $K=18$ for the six-comparator-by-three-regime family. For ordered marginal values $p_{(1)}\leq\cdots\leq p_{(K)}$, the adjusted value is $\widetilde p_{(k)}=\min\{1,\max_{\ell\leq k}(K-\ell+1)p_{(\ell)}\}$. At $\widetilde p<0.05$, negative/positive estimated log ratios support lower Neuro-Dist/control error; no statistically significant difference is detected in the other comparisons, which does not establish equivalence. Holm requires no additional dependence assumption; familywise control remains approximate because BCa inference uses a finite sample of independent seed blocks.

\begin{table}[htbp]
\caption{\textbf{Adjusted aggregate significance.} Two-sided BCa confidence-curve $p$-values with Holm correction across the same 18 aggregate comparisons as Table~\ref{tab:accuracy-statistics-aggregate}. Values preceded by $\leq$ are conservative upper bounds resulting from the marginal numerical floor.}
\label{tab:accuracy-statistics-pvalues}
\begin{center}
\begingroup
\setlength{\tabcolsep}{3pt}
\begin{tabular}{@{}lccc@{}}
\toprule
Comparator & Trained targets & Student-held-out & New targets \\
\midrule
Task-only & $\leq 1.8\times10^{-4}$ & $0.0284$ & $0.0284$ \\
Ring & $\leq 1.8\times10^{-4}$ & $0.0040$ & $9\times10^{-4}$ \\
Behavior & $0.0046$ & $0.0931$ & $0.0284$ \\
Shuffled & $\leq 1.8\times10^{-4}$ & $\leq 1.8\times10^{-4}$ & $\leq 1.8\times10^{-4}$ \\
Gaussian & $\leq 1.8\times10^{-4}$ & $\leq 1.8\times10^{-4}$ & $\leq 1.8\times10^{-4}$ \\
Poisson & $\leq 1.8\times10^{-4}$ & $0.0046$ & $0.0046$ \\
\bottomrule
\end{tabular}

\endgroup
\end{center}
\end{table}

\paragraph{Numerical checks and inferential scope.}
The four-million-draw budget was fixed before evaluating the new-dose new-target outcomes. Numerical checks assess median ties and gaps, bootstrap-tail resolution, interval endpoints, and Holm-corrected decisions in the full bank and its two equal halves, without fallback, filtering, or significance-dependent resampling. Inference concerns seed variability in three \emph{fixed} animal datasets, conditional on the frozen selection half and dose map.

\subsection{Results}

\paragraph{Aggregate comparisons.}
Figure~\ref{fig:accuracy-statistics} and Tables~\ref{tab:accuracy-statistics-aggregate}--\ref{tab:accuracy-statistics-pvalues} summarize the seed-split validation. Holm-corrected comparisons support lower Neuro-Dist median MSE against all six comparators on trained targets (geometric-average reductions 26.9\%--51.1\%), five on student-held-out targets (29.1\%--43.5\%), and all six on new directions (16.7\%--44.5\%). Task-only reductions are 30.7\% on student-held-out targets and 27.5\% on new directions (both adjusted $p=0.0284$). The new-direction reduction against Behavior is 16.7\% (adjusted $p=0.0284$); no statistically significant difference is detected against Behavior on student-held-out targets (15.2\%, adjusted $p=0.0931$). Thus, 17 of 18 aggregate comparisons favor Neuro-Dist, with none favoring a control.

\paragraph{Numerical resolution.}
All 18 aggregate Holm decisions agree across the full bootstrap bank and both halves. No cell has a central median tie.


\end{document}